\documentclass[runningheads]{llncs}
\PassOptionsToPackage{obeyspaces}{url}
\usepackage{graphicx}
\usepackage{float, enumerate, xcolor, algorithm, algorithmic, tabto, balance, amsmath, graphicx, epstopdf, hyperref, multirow, dblfloatfix, textcomp, listings, eurosym, lstautogobble, multirow, url, tikz} 
\usepackage{amsfonts}

\usepackage{amssymb}
\usepackage{booktabs}
\usepackage{url}
\usepackage{amssymb}
\usepackage[numbers,sort&compress]{natbib}
\usepackage[flushleft]{threeparttable}
\usepackage{subfigure}
\usetikzlibrary{arrows,backgrounds,positioning}
\usepackage{chngcntr}
\usepackage[overload]{empheq}
\usepackage{soul}
\usepackage[multiple]{footmisc} 
\usepackage[most]{tcolorbox} 
\usepackage[title]{appendix}
\usepackage{subfigure}

\usepackage{soul, listings, color}

\definecolor{dkgreen}{rgb}{0,0.6,0}
\definecolor{gray}{rgb}{0.5,0.5,0.5}
\definecolor{mauve}{rgb}{0.58,0,0.82}

\usepackage{chngcntr}
\AtBeginDocument{\counterwithin{lstlisting}{section}}

\newcommand\sbullet[1][.5]{\mathbin{\vcenter{\hbox{\scalebox{#1}{$\bullet$}}}}}

\newcommand{\riccardo}[1]{\textcolor{black}{#1}}

\newcommand{\ankit}[1]{\textcolor{black}{#1}}

\newcommand{\removed}[1]{\textcolor{red}{
}}

\usepackage{array}
\newcolumntype{L}{>{\arraybackslash}m{.45\columnwidth}|}

\usepackage{url}

\usepackage{pifont}
\usetikzlibrary{arrows,automata}

\graphicspath{ {./images/} }

\begin{document}
	\title{BLEWhisperer: Exploiting BLE Advertisements for Data Exfiltration}
	\titlerunning{BLEWhisperer}
	\author{Ankit~Gangwal\inst{1}
		\and
		Shubham~Singh\inst{1}
		\and
		Riccardo~Spolaor\inst{2,*}
		\and
		Abhijeet~Srivastava\inst{1}
	}
	\authorrunning{A.~Gangwal et al.}
	\institute{International Institute of Information Technology, Hyderabad, India \\
		\email{gangwal@iiit.ac.in}, \\
		\email{\{shubham.singh, abhijeet.srivastava\}@students.iiit.ac.in} 
		\and
		Shandong University, Qingdao Campus, China\\
		\email{rspolaor@sdu.edu.cn}\\
		* Corresponding author\\
		\vspace{-.75em}}
	
	\maketitle
	\begin{abstract}
		Bluetooth technology has enabled short-range wireless communication for billions of devices. Bluetooth Low-Energy (BLE) variant aims at improving power consumption on battery-constrained devices.  BLE-enabled devices broadcast information (e.g., as beacons) to nearby devices via advertisements. Unfortunately, such functionality can become a double-edged sword at the hands of attackers.
		\par
		In this paper, we primarily show how an attacker can exploit BLE advertisements to exfiltrate information from BLE-enable devices. In particular, our attack establishes a communication medium between two devices without requiring any prior authentication or pairing. We develop a proof-of-concept attack framework on the Android ecosystem and assess its performance via a thorough set of experiments. Our results indicate that such an exfiltration attack is indeed possible though with a limited data rate. Nevertheless, we also demonstrate potential use cases and enhancements to our attack that can further its severeness. Finally, we discuss possible countermeasures to prevent such an attack.		
	\end{abstract}
	\section{Introduction}
	Bluetooth is a pervasive wireless technology that is widely used for building Personal Area Network~(PAN). 
	Bluetooth open standard~\cite{BL5.2} specifies two paradigms: Bluetooth Classic (BT) and Bluetooth Low Energy (BLE). While BT is suitable for high-throughput communication, BLE is designed for low-power communication. 
	BLE protocol enables two devices to exchange data with one device acting~as~a client and another one as a server. 
	According to the current specifications~\cite{BL5.2}, Bluetooth 5.2 quadruples the transmission range (LE coded eight symbols per bit) compared to the last generation (i.e., Bluetooth 4.2)~\cite{BL4.2}. Studies~\cite{BLMarketUpdate, ABIResearch} estimate that manufacturers will ship nearly 6.3~Billion Bluetooth-enabled devices by 2025, among which 6~Billion devices will support BLE. On another side, mobile devices have adopted the Bluetooth technology to offer wireless connectivity among devices and with other peripherals, such as headphones. Among mobile Operating Systems~(OS), Android covers the largest share of the market~\cite{Statista}. As of Q3~2021, Android's share of mobile OS worldwide was 72.84\%~\cite{Statista} with 17.5\% devices running on Android~11, 35.9\% devices using Android~10, and 46.6\% devices still operating with Android~9 or below~\cite{StatCounter}. Generally, mobile devices are shipped with few apps pre-installed, and end-users can install different apps to enhance/customize user-experience. Depending upon granted permissions, such apps can also use the device's Bluetooth radio.
	\par
	Bluetooth technology has evolved greatly over time and continuous efforts have been made to make its entire stack secure. Nonetheless, about 75~Bluetooth-related CVEs~\cite{cve} were reported in the year 2021 alone. BLE advertisements are no exception. Disclosing a device's presence via advertising can lead to privacy and security attacks; an adversary can monitor advertisements to gather information about the advertising BLE device~\cite{x1, x2}. The core BLE specification stipulates some privacy provisions (in particular, whitelisting and address randomization) to tackle these threats. Device whitelisting focuses on device pairing while address randomization hinders others from tracking a device over time.
	\par
	\textit{Motivation:} Each connection in BLE communication starts its lifetime by advertising primary information. In particular, BLE advertisements enable devices to exchange their capabilities, characteristics, etc. even before pairing happens. However, such information exchange mechanism lacks proper security measures to prevent its misuse. So, it is necessary to investigate to what extents an attacker can exploit such functionality and its consequent security risks.
	
	\par
	In this paper, we investigate the possibility of exploiting BLE advertisements as a communication channel between attacker and target device; using which an attacker may issue commands to perform some tasks, deliver arbitrary payload when other channels (e.g., WiFi and data) are restrained, bypass address randomization defense, etc. Specifically, our attack utilizes \textit{service data type} of BLE advertisements that can carry arbitrary values in its Service Data field, which makes it suitable to transmit custom data. Furthermore, we employ non-connectable BLE advertisements to enable our attack even if the victim device is connected/paired to another Bluetooth device. \ankit{Our attack prototype targets Android~OS to cover the majority of mobile devices. We begin with BLE legacy advertisements, which are supported by both Bluetooth~4.2~(adopted in 2014~\cite{BL4.2}) and the latest Bluetooth~5 family} \ankit{~\cite{BL5.2}. We also demonstrate our attack leveraging extended advertisements of Bluetooth~5, which further increases its data transfer capabilities.} Our attack requires the attacker to be in the Bluetooth range of a victim and that the victim has installed our app, which we call \textit{victim's app}. To communicate, the attacker and \textit{victim's app} use BLE advertisements.
\par
\textit{Contribution:} The major contributions of our work are as follows:
\begin{enumerate}
	\item We primarily demonstrate how an adversary can exploit BLE advertisements as a communication channel between attacker and target device. We propose a data exfiltration attack via BLE advertisements that does not require authentication or pairing. We fully implemented all the components required for such an attack. To prevent misuse, source code is available on request.
	\item \ankit{To thoroughly assess our proposed attack, we designed two experiments that we conduct on five smartphones for both BLE legacy and extended advertisements. Our results show that such an attack indeed poses a threat.}
	\item We also discuss further enhancements, key use cases, and possible countermeasures of our proposed attack.
\end{enumerate}	
\par
\textit{Organization:} The remainder of this paper is organized as follows. Section~\ref{section:background} summarizes the background for BLE and related works. We explain our threat model and the core idea of our attack in Section~\ref{section:system_architecture}. Section{~\ref{sec:PoCimplementation}} gives the details of our proof-of-concept implementation. Section~\ref{section:evaluation} reports our experimental evaluations. \ankit{Section~\ref{section:discussion} presents salient add-ons, use cases, potential limitations, and countermeasures for our attack.} Finally, Section~\ref{section:conclusion} concludes the paper.

\section{Background}
\label{section:background}
In this section, we present the primer for BLE advertising in Section{~\ref{Section:BLE}} and a summary of related works in Section{~\ref{section:related_work}}.
\subsection{Bluetooth Low Energy (BLE)}
\label{Section:BLE}
BLE~\cite{BLTech} is a low-power wireless technology typically used for short-distance communication. Both BLE and BT operate in the same 2.4 GHz ISM band. 
\par
\textit{Advertising:} BLE advertisements are used to notify nearby devices of the availability to make a connection. Here, a Bluetooth device can assume two major roles:  advertiser~(as peripheral or broadcaster) and scanner~(as central or observer). Advertisers create and transmit the advertisements while scanners receive these advertisements. BLE has 40 RF channels, where 3 channels (i.e., channels 37, 38, and 39) are used for advertisements. In BLE, the time interval between advertisements has a fixed interval as well as a random delay~\cite{BL4.2}.  Legacy Protocol Data Unit (PDU) advertisements (i.e., \texttt{ADV\_DIRECT\_IND, ADV\_IND, ADV\_NONCONN\_IND, ADV\_SCAN\_IND}) are available for all Bluetooth versions, have backward compatibility with older versions, and are used on the Primary advertising channels. Extended PDU advertisements (i.e., \texttt{ADV\_EXT\_IND, AUX\_ADV\_IND, AUX\_SCAN\_IND, AUX\_CHAIN\_IND}), introduced in Bluetooth~5.0, enable advertising on Secondary advertising channels (in addition to Primary advertising channels) to increase advertising data capacity.
\par
\begin{figure}[H]
	\centering
	{\includegraphics[trim = 0mm 0mm 0mm 0mm, clip, width=0.45\columnwidth]{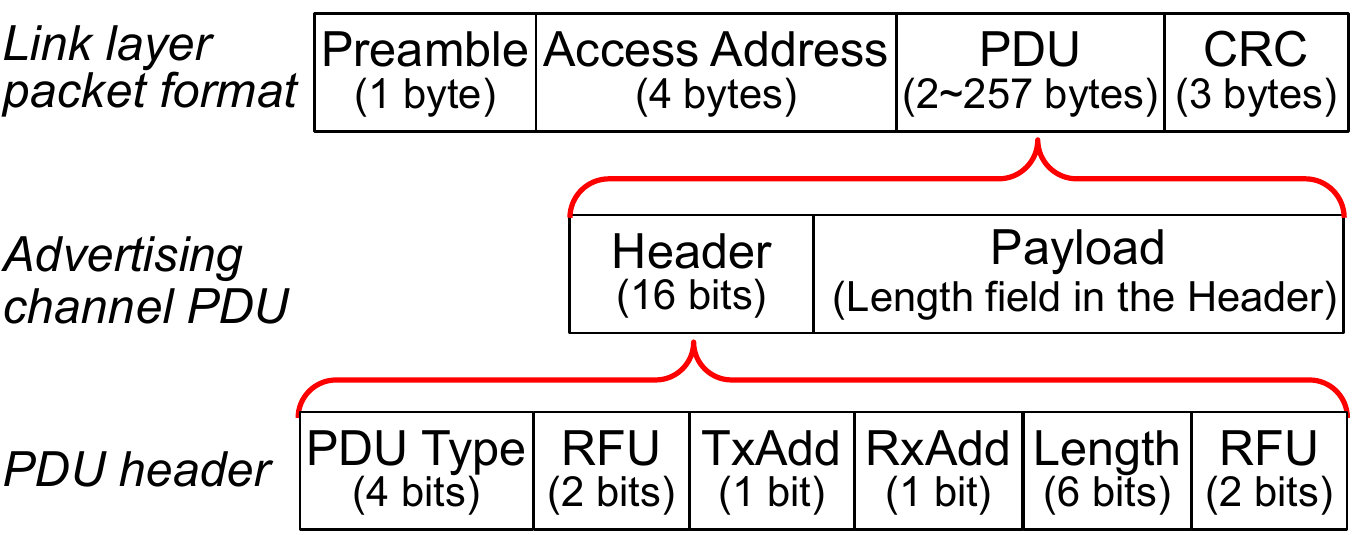}}
	\vspace{-.5em}
	\caption{BLE packet structure~\cite{BL4.2}.}
	\label{fig:PDU1}
	\vspace{-1em}
\end{figure}

\textit{Packet format:} The core Bluetooth specification document~\cite{BL4.2} defines the link layer packet in BLE with preamble, access address, PDU, and CRC. PDU for advertising channel (called advertising channel PDU) includes a 2-byte header and a variable payload (from 6 to 37 bytes), whose  actual length is defined by the 6-bit Length field of advertising channel PDU header~(cf. \figurename{~\ref{fig:PDU1}}). Since BLE supports a number of standard advertisement data types (e.g., \textit{manufacturer specific data, service solicitation, service data, LE supported features}) that can be sent in an advertisement, the content of advertising channel PDU payload depends on the chosen advertisement data types. 
\par
\textit{Universally Unique Identifier~(UUID):} A client searches for services based on some desired characteristics. A BLE profile can offer one or more services, and each service can have one or more characteristics. Each service distinguishes itself from other services using a unique 16-bytes hexadecimal ID, called UUID. While standard services can use 2- or 4-bytes UUID to make room for more data in advertisements, custom services require a full 16-bytes UUID. 
\vspace{-.5em}
\subsection{Related works}
\label{section:related_work}
Researchers have been working towards enhancing the security of the Bluetooth technology by exploring possible exploits and attacks. In what follows, we report the main works related to our paper. BIAS~\cite{bias} bypasses the authentication step to impersonate an already paired benign Bluetooth device.
Similarly, BLESA~\cite{blesa} exploits a BLE protocol vulnerability to inject malicious data when a smartphone reconnects to a paired device.
BLURtooth~\cite{blurtooth} proposes cross-transport attacks on active session and leads to device impersonation, malicious session establishment, and manipulation of Bluetooth traffic.
BlueDoor~\cite{bluedoor} targets connected BLE devices and mimics a low-capacity device to undermine the process of key negotiation and authentication. 
LIGHTBLUE~\cite{lightblue} is a framework for performing automatic profile-aware debloating of Bluetooth application stack. However, LIGHTBLUE is not designed for general users since it requires advanced technical skills, such as phone rooting, installing modified firmware, etc.
BadBluetooth~\cite{badbluetooth} attack can steal information, sniff network traffic, and inject voice command on a device with compromised firmware.
With the help of specialized hardware and software components, BLE-guardian framework~\cite{fawaz2016protecting} jams the advertising channel to hide a device's presence from curious adversaries.
BlueShield~\cite{blueshield} presents a monitoring framework that detects spoofed BLE advertisements against a stationary BLE network in indoor environments. Armis demonstrated an airborne attack vector called 
BlueBorne~\cite{blueborne}. In the context of smartphones, BlueBorne CVEs affect devices running upto Android~8.0 and iOS~10. \ankit{Singh et al.~\cite{singh2010evaluating} present mobile phone-based botnets that utilize Bluetooth connection alongside cellular channel for communication. The Bluetooth standards and connection establishment mechanisms have evolved since the time of the study and become more complex and restrictive.}
\par
To summarize, existing works target already paired devices~\cite{bias, blesa, blurtooth, bluedoor}, require a compromised firmware~\cite{lightblue, badbluetooth}, specialized hardware components~\cite{fawaz2016protecting}, or only work under specific settings~\cite{blueshield, singh2010evaluating}. To the best of our knowledge, we are the first to investigate the misuse of BLE advertisements to create a communication channel and its security implications.

\section{System architecture}
\label{section:system_architecture}	
In this section, we present the system's architecture for our attack.
Section{~\ref{section:threat_model}} elaborates the threat model, Section{~\ref{section:thecoreIdea}} explains the core idea our work, and Section{~\ref{section:process}} discusses different phases of our attack.

\subsection{Threat model}
\label{section:threat_model}
Our attack relies on two practical assumptions:~(i) the victim eventually comes in the Bluetooth range of the attacker and (ii)~the victim installs our benign-looking app, i.e., \textit{victim's app}, which can come from the genuine application store. 
Overall, the attacker shares a context with the victim.

\textit{Victim's app} requires following permissions~\cite{AndroidManifestPermission}: (i)~Bluetooth permissions~(i.e., \texttt{BLUETOOTH} and~\texttt{BLUETOOTH\_ADMIN}) to administer/toggle Bluetooth radio, and (ii)~location permission~(i.e., \texttt{ACCESS\_FINE\_LOCATION}) to scan Bluetooth advertisements. Both the Bluetooth-related permission are normal\footnote{Normal permissions are granted without explicit user consent/interaction.} while the location permission is designed to be dangerous\footnote{Dangerous permissions are granted only if user \ankit{explictily consents} to it.}. Most of today's apps (navigation, taxi, food delivery, contact tracing, etc.) rely on location services to provide their service or to verify user's location. Therefore, \textit{victim's app} can come in a variety of forms to request the location permission. Depending on the OS version, location service may be required to turned on~(cf. Section{~\ref{section:and10+}}).
\riccardo{Some scenarios where apps verify user's location include attendance app for students in a classroom, sign-on app for employees in an office, boarding pass app for airline passengers, and apps for public events (e.g., conference, concerts, museum).}
\par
\riccardo{As the majority of malware rely on an Internet connection to steal user data, a network-based Intrusion Detection System~(IDS) can identify such exfiltration attempts and trigger an alert. Hence, data exfiltration via BLE may be a viable solution when the Internet connection is monitored, restricted, or unavailable~(e.g., in an airplane, air-gapped networks).}

\vspace{-.75em}
\subsection{The core idea}
\label{section:thecoreIdea}
\begin{figure}[H]
	\vspace{-1em}
	\centering
	{\includegraphics[trim = 0mm 0mm 0mm 0mm, clip, width=0.3\columnwidth]{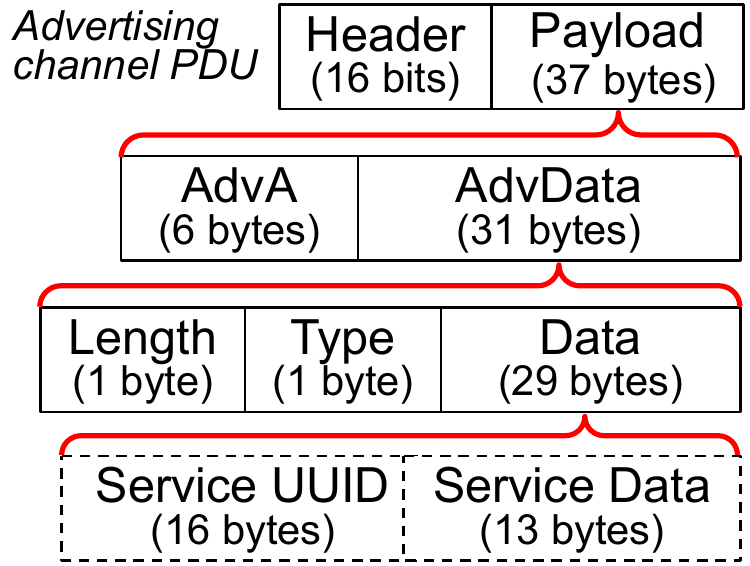}}
	\vspace{-.5em}
	\caption{Advertising channel PDU for \textit{service data type}~\cite{BL4.2}.}
	\label{fig:BLE4PktFrmt}
	\vspace{-1.75em}
\end{figure}

\vspace{-.6em}
Among various BLE advertisement data types, \textit{service data type} allows us to set arbitrary values in its Service Data field; which makes it suitable to transmit custom messages. \figurename{~\ref{fig:BLE4PktFrmt}} shows the format of advertising channel PDU for the \textit{service data type} (cf.~\figurename{~\ref{fig:PDU1}} for 2-byte header field structure) \ankit{in legacy advertisements.} Here, advertising channel PDU payload contains 6-bytes AdvA field (i.e., advertiser's address) and upto 31-bytes AdvData field (i.e., advertised data). AdvData field contains 1-byte Length field, 1-byte Type field, and 29-bytes Data field. Data field contains 16-bytes Service UUID and 13-bytes Service Data.
The core idea of our attack is to leverage the Service Data field to transport custom message payloads. Our attacker advertises \textit{service data type} with an attacker-chosen fixed Service~UUID~(hardcoded in \textit{victim's app}) and Service Data field carrying commands encoded to bytes. It is worth mentioning that the requested Bluetooth permissions enable \textit{victim's app} to toggle Bluetooth on/off without user's intervention. By using non-connectable BLE advertisements, our attack can work even if the victim device is connected to other Bluetooth devices.
\par
Our attack setting involves an attacker device and one or more victim devices. A victim device is an Android device with our \textit{victim's app} installed. The attacker device is a Windows laptop with a Bluetooth interface. Both devices act as BLE advertisers and scanners from time to time. Attacker sends the commands as advertisement broadcast with a particular Service UUID. \textit{Victim's app} scans for advertisements; it responds when attacker's UUID is matched. Conversely, to transmit data from victim to attacker, \textit{victim's app} advertises data in the same manner (i.e., in Service Data field of \textit{service data type}), but with victim-specific UUID in the Service UUID field.
\par
\textit{UUIDs and their roles:}	UUID \ankit{plays} a crucial role in our attack. Hence, it is important to understand the roles of different UUIDs. $UUID_{A}$ is an attacker-chosen fixed UUID hardcoded in \textit{victim's app}. $UUID_{A}$  is what \textit{victim's app} listens for. $UUID_{V}$ is victim-specific UUID that is generated by victim device's OS; it may change across different connections. $ID_{V}$ is a victim-specific identifier generated by \textit{victim's app}, and it is permanent for a victim device. We map  $UUID_{V}$ to $ID_{V}$ to identify/track the same victim across different connections. However, the first time a victim's device responds, its $ID_{V}$ is unknown to the attacker. Therefore, \textit{victim's app} uses a special pattern ``$UUID_{V}$ , \texttt{0x000000} $\|$ ~$ID_{V}$'', i.e., it sends \texttt{0x000000} concatenated with $ID_{V}$ (in Service Data field) from its current $UUID_{V}$ (in Service UUID field) to signify to attacker that after \texttt{0x000000}~(a pre-decided value) is an $ID_{V}$, and the attacker maps the two values.
\par
\begin{figure}[!htbp]
	\vspace{-.5em}
	\centering
	\includegraphics[trim = 0mm 0mm 0mm 0mm, clip, width=0.55\columnwidth]{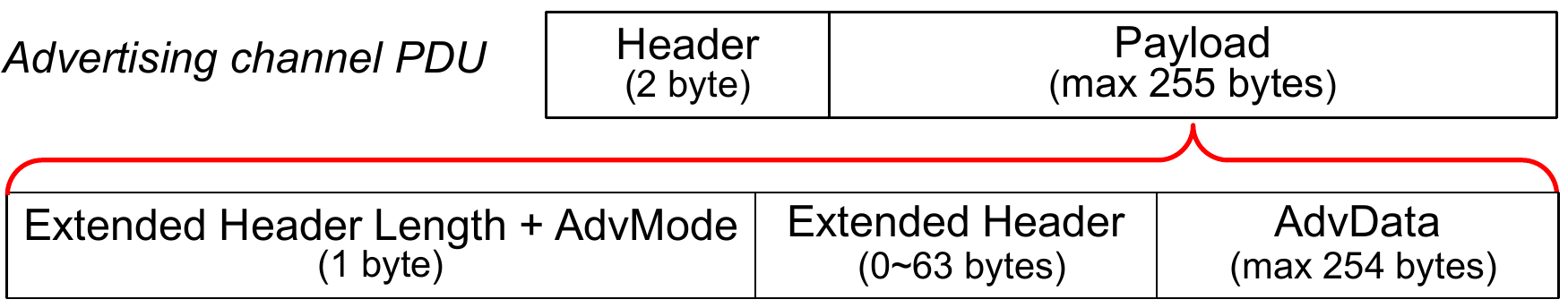}
	\vspace{-.5em}
	\caption{\ankit{Advertising channel PDU for BLE 5.2.}}
	\label{fig:BLE5PktFrmt}
	\vspace{-1em}
\end{figure}

\ankit{\textit{Increased impact with BLE 5:} Along with longer transmission range and higher data throughput, Bluetooth~5 also offers advertising extensions. Instead of advertising only on the three advertising channels (i.e., channels 37, 38, and 39), BLE 5 allows to chain together advertisements and utilize other 37 RF channels for advertisements. Moreover, advertising channel PDU payload for BLE 5 can hold up to 254 bytes of AdvData~\cite{BL5.2}~(cf. \figurename{~\ref{fig:BLE5PktFrmt}}), which is about 8 times of 31 bytes of AdvData in BLE4~\cite{BL4.2}.}
\par
\ankit{In our prototype, we considered both the legacy and extended advertisements. The former is compatible with the widest range of mobile devices, and the latter is becoming increasingly common among newer devices.}
\vspace{-.45em}
\subsection{Attack phases}
\label{section:process}
\begin{figure}[ht]
	\vspace{-.5em}
	\centering
	\subfigure[Discovery \& Selection.] 
	{\label{fig:discoveryandcontact}
		\includegraphics[width=0.38\columnwidth, height=10cm]{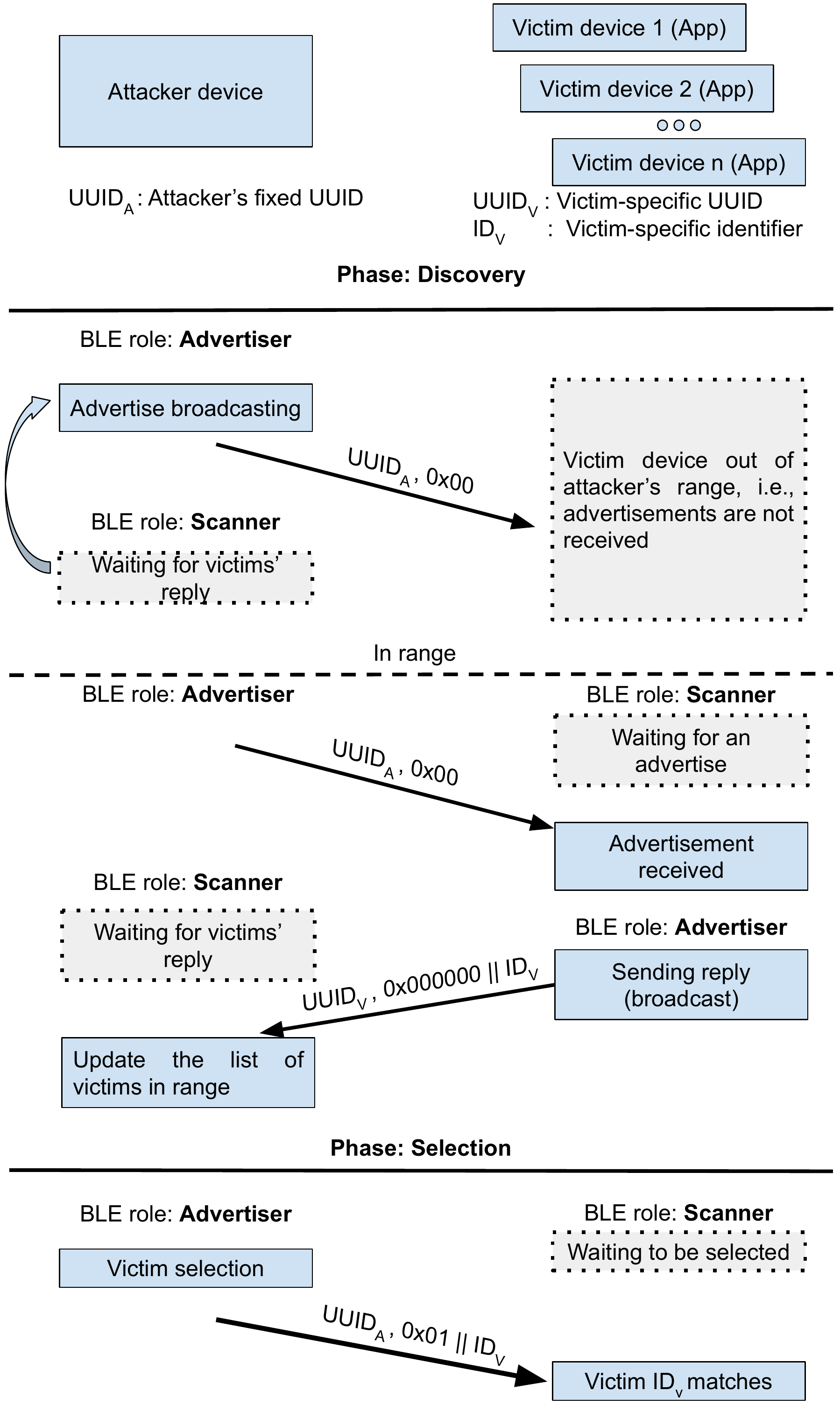}
	}\hspace{1mm}
	\subfigure[Transfer, Validation, \& Recovery.] 
	{\label{fig:transferandrecovery}
		\includegraphics[width=0.48\columnwidth, height=10cm]{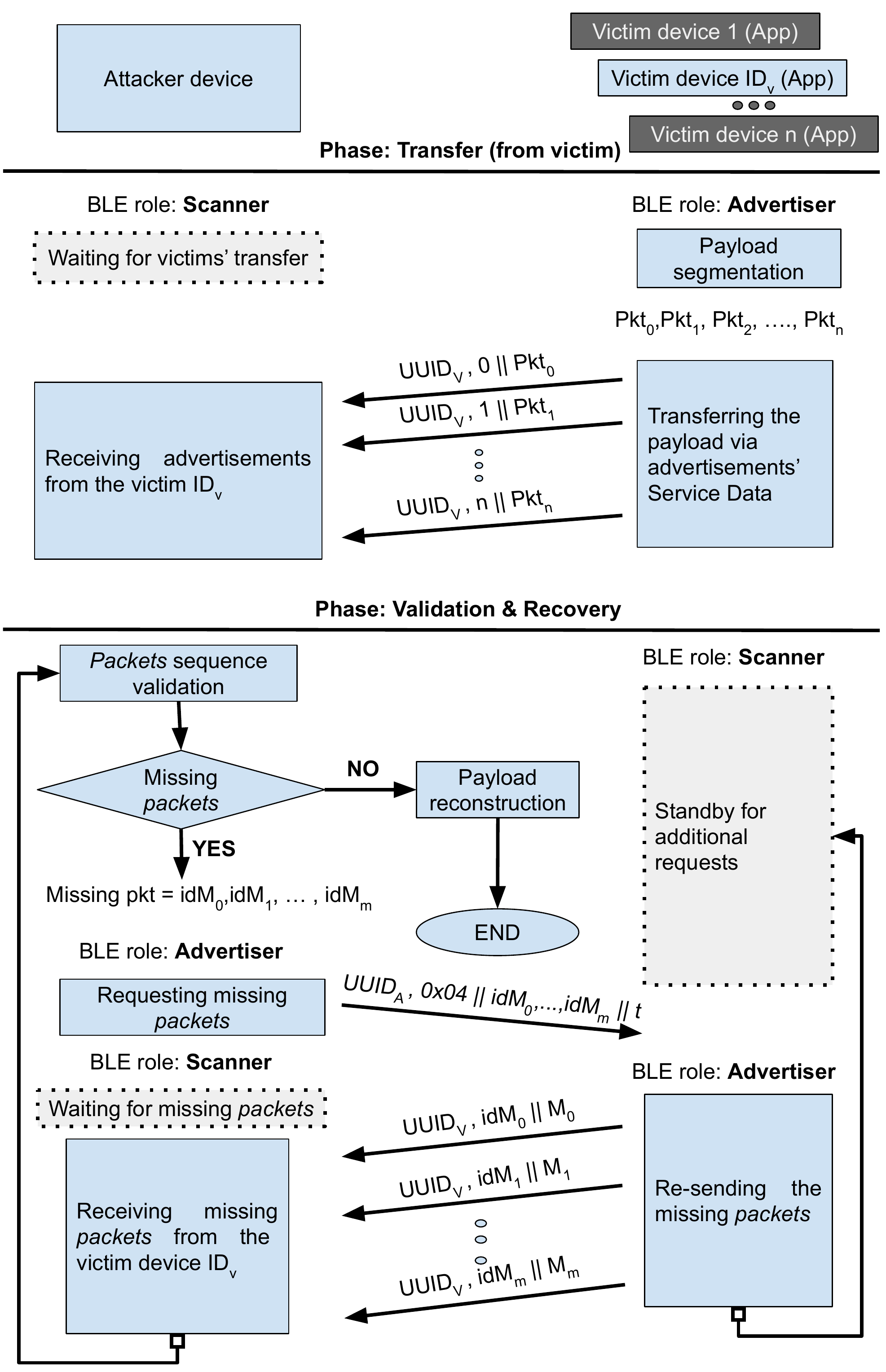}}
	\vspace{-.5em}
	\caption{Phases of our attack}
	\label{fig:attackphases}
	\vspace{-1.2em}
\end{figure}

In the default state, a victim device scans for BLE advertisements with attacker's UUID~(i.e., $UUID_{A}$) to receive instructions. \figurename~\ref{fig:attackphases} shows different phases of our attack. We now elaborate each phase in detail. 
\par
\textit{$\sbullet[.75]$ Discovery phase:} In the first phase of our attack (i.e., Discovery phase in \figurename~\ref{fig:discoveryandcontact}), the attacker device starts with advertiser mode, where it broadcasts an advertisement containing $UUID_{A}$ and our custom discovery command as byte \texttt{0x00}. After sending this discovery advertisement, the attacker switches to scanner mode to scan for any reply advertisement from victim devices in the transmission range. In case of no response is received within an interval of time, the attacker switches back to the advertiser mode and repeats the process (i.e., discovery broadcasting and scanning for reply). Victim devices in the transmission range would receive the discovery advertisement. Subsequently, the victim devices switch to advertiser mode, broadcast a reply advertisement containing $UUID_{V}$, a special pattern (cf. Section{~\ref{section:thecoreIdea}}), and $ID_{V}$; switch again to scanner mode and wait to be selected by the attacker device. The attacker maintains a list of victim devices currently in range (using the mapping of $UUID_{V}$ to $ID_{V}$). When one or more advertisement replies are received, the attacker updates such a list by adding the relevant victim devices' information. At the same time, when a victim device does not reply to a discovery advertisement, then the attacker would consider that victim device as unreachable (i.e., out of range).
\par
\textit{$\sbullet[.75]$ Selection phase:} To select a victim device (i.e., Selection phase in \figurename~\ref{fig:discoveryandcontact}), the attacker device switches to advertiser mode, broadcasts selection command as byte \texttt{0x01} and target victim's $ID_{V}$. After receiving this selection announcement, victims check if they are selected by comparing their $ID_{V}$, and only the selected one is involved in subsequent phase. Meanwhile the others victim devices wait for the next discovery advertisement~(i.e., a new Discovery phase). After sending the selection announcement, the attacker instructs the selected victim to start data transmission and switches to scanner mode.
\par	
\textit{$\sbullet[.75]$ Transfer phase:} The selected victim device segments the payload to be exfiltrated into $n$ enumerated segments of maximum size $z$ bytes, then it switches to advertiser mode. Each advertisement from the victim has its $UUID_{V}$ in Service UUID field while Service Data field contains segment's number and segment's content~(i.e., Transfer phase in  \figurename~\ref{fig:transferandrecovery}). In this phase, the victim device broadcasts all the enumerated segments in a sequence. At the same time, attacker receives and saves segments into memory. At the end of transmission, victim device switches to scanner mode and wait for further instructions. Henceforth, we refer as \textit{packets} to the advertisements sent by \textit{victim's app} that include a segment of payload to be exfiltrated.
\par
\textit{$\sbullet[.75]$ Validation \& Recovery phases:} Due to possible transmission problems, some \textit{packets} may not be well received by the attacker. For this reason, the attacker runs Validation \& Recovery phases~(shown in \figurename~\ref{fig:transferandrecovery}). The attacker verifies if all the \textit{packets} have been received correctly~(i.e., validation). If \textit{packets} are missing, the attacker device in advertiser mode requests the victim device to send the missing \textit{packets}~(i.e., recovery) by sending a sequence with the missing \textit{packets'} numbers; then attacker waits in scanner mode. Once the victim device receives such a request, it switches to advertiser mode to send the missing \textit{packets} and waits for further instructions in scanner mode. The attacker verifies correct reception of all \textit{packets}. If any \textit{packet} is still missing, the attacker repeats the recovery and validation steps until all \textit{packets} have been received correctly. Finally, the attacker reconstructs the payload from the entire \textit{packet} sequence. 

\vspace{-.75em}
\section{Proof-of-concept implementation of the attack}
\label{sec:PoCimplementation}
\vspace{-.25em}
To carry out our attack, we design and implement a proof-of-concept framework.
In this section, we describe the implementation of our framework's components at the attacker side (in Section{~\ref{section:advertiser}}) and victim side (in Section~\ref{section:victim}).

\subsection{Attacker side: $AT_{Advt}$ and $AT_{Scan}$}
\label{section:advertiser}
On the attacker side, we developed two applications: $AT_{Advt}$ and $AT_{Scan}$. $AT_{Advt}$ is in charge of broadcasting advertisements and acts as a controller for the data transmission by \textit{victim's app}. $AT_{Scan}$ acts as a receiver and it continuously listens for advertisements from a \textit{victim's app} (filtered by $UUID_{V}$).
We implemented $AT_{Advt}$ in C\# and $AT_{Scan}$ in Python 3.10.1 using the Bleak libraries~\cite{BLEAK}. 
All advertisements from $AT_{Advt}$ contain $UUID_{A}$ as the Service UUID and a command (with its arguments) encoded as bytes in Service Data field. $AT_{Advt}$ can issue four types of commands: \textit{victim's app}'s discovery, target selection, start/stop transmission, and \textit{packet} retransmission request. Next, we describe the details of these commands from both $AT_{Advt}$ and $AT_{Scan}$ points of view.

\textit{$\sbullet[.75]$ Victim's apps' discovery (command byte }\texttt{0x00}\textit{):}
$AT_{Advt}$ sends an advertisement to discover the presence of all \textit{victim's app}(s) in range. Such advertisement includes our discovery command as byte~\texttt{0x00} in Service Data field. In the meantime, $AT_{Scan}$ monitors reply advertisements from in-range victim devices and updates the mapping of $UUID_{V}$ to $ID_{V}$. In particular, $AT_{Scan}$ filters such replies via an identifier (i.e., fixed starting bytes \texttt{0x000000}) in Service Data field.

\textit{$\sbullet[.75]$ Target selection (command byte }\texttt{0x01}\textit{):}
The attacker can select a particular victim from the list of currently in-range victim devices. To do so, $AT_{Advt}$ sends an advertisement that includes this command as byte \texttt{0x01} followed by $ID_{V}$ of target  victim device in Service Data field. From now on, only the target \textit{victim's app} would respond to further commands.

\textit{$\sbullet[.75]$ Start and stop transmission (commands bytes }\texttt{0x02}\textit{ and }\texttt{0x03}\textit{, respectively):}
$AT_{Advt}$ sends an advertisement that includes the command to start payload transmission (i.e., byte \texttt{0x02}) or to stop an ongoing one~(i.e., byte \texttt{0x03}) in Service Data field. In particular, the start transmission command also sends along parameter $t$, which specifies victim's data transmission speed in terms of time interval between its successive advertisements. $AT_{Scan}$ would collect advertisements coming from target's $UUID_{V}$, which is mapped to $ID_{V}$.

\textit{$\sbullet[.75]$ Retransmission request (command byte }\texttt{0x04}\textit{):}
$AT_{Advt}$ can request retransmission of one or more missing \textit{packets} from \textit{victim's app}. Since \textit{victim's app} includes corresponding segment's number in a \textit{packet}, $AT_{Advt}$ can issue a retransmission request with command byte \texttt{0x04} followed by segment numbers of missing \textit{packets} and parameter \textit{t} in Service Data field. Similar to the start/stop transmission command, $AT_{Scan}$ would also collect retransmitted advertisements.
\par
\lstlistingname~\ref{lst:advertiserModel} in \appendixname{~\ref{appendix:codeSnippets}} shows advertisement manipulation by $AT_{Advt}$.
\vspace{-.75em}
\subsection{Victim side: \textit{Victim's app}}
\label{section:victim}
\vspace{-.25em}
On the victim side, we developed  \textit{victim's app} running on an BLE-enabled Android device. We implemented this app using Android Studio Version 2020.3.1. We built our \textit{victim's app} using SDKv30 and SDK minVer26. $UUID_{A}$ is hard-coded (a standard practice) in \textit{victim's app}, so it can recognize advertisements from the attacker. \textit{Victim's app} includes both a scanner mode~(to listen to $AT_{Advt}$ commands) and an advertiser mode~(to send advertisements). 
We report the configuration codes for the scanner and advertiser modes of  \textit{victim's app} in Listings~\ref{bluetoothLeScannerSettings} and~\ref{setAdvertiseSettingsAdvertiseCallback}, respectively in \appendixname{~\ref{appendix:codeSnippets}}.
Now, we describe in detail \textit{victim's app} actions according to $AT_{Advt}$ commands.

\textit{$\sbullet[.75]$ Response to discovery command:} 
Upon receiving discovery command, \textit{victim's app} builds and sends a response advertisement, which contains $UUID_{V}$~(OS enforced, can change overtime) as the Service UUID and \texttt{0x000000} followed by its $ID_{V}$ in the Service Data field.

\textit{$\sbullet[.75]$ Selected as target:} Upon receiving a target selection command (that contains target's $ID_{V}$), a \textit{victim's app} matches its own $ID_{V}$ against the one in the advertisement. If it matches, then this \textit{victim's app} expects further commands from the attacker. From now on, only the target \textit{victim's app} responds to further commands. All other victim devices wait for a new Discovery phase.

\textit{$\sbullet[.75]$ Data transmission:} With a start command from $AT_{Advt}$, the attacker tells the target to transmit payload through \textit{packets}. Since the data to be exfiltrated has to be segmented over multiple \textit{packets}, we store the segment's number in the first byte of the Service Data field of each \textit{packet}. The segment's number helps to identify any duplicate as well as lost \textit{packets} to be retransmitted. Since the Service Data field can contain at most 13 bytes in total, each \textit{packet} consists of 1 byte of segment's number and 12 bytes of segment's data. 
Moreover, \textit{victim's app} also scans (i.e., bidirectional radio) for advertisements from $AT_{Advt}$ with a command to stop the transmission. 

\textit{$\sbullet[.75]$ Retransmission request:} 
Responding to a retransmission request, \textit{victim's app} creates and sends missing \textit{packets} identified by segment numbers. 
\par
To reduce the number of explicit retransmission requests in recovery phase, we designed \textit{victim's app} to transmit the entire sequence of payload \textit{packets} a certain number of consecutive times defined by parameter~$R$; i.e., \textit{victim's app} transmits all the payload \textit{packets} and repeats the process $R$ times. Thus,  $AT_{Scan}$ can receive a specific \textit{packet}~$R$ times at most. Alternatively, parameter~$T$ defines the timeout until which \textit{victim's app} keeps on sending all the payload \textit{packets}, i.e., $R=\infty$ till $T$. $AT_{Advt}$ can issue a stop transmission command when required.
\vspace{-.5em}
\section{Experimental evaluation}
\label{section:evaluation}
\vspace{-.5em}
We describe our hardware setup and experimental method in Section{~\ref{section:hw}}. \ankit{We report the analysis of our results for BLE legacy and extended advertisements in Section{~\ref{section:resultsLegacy}} and Section{~\ref{section:resultsExtended}}, respectively.}
\vspace{-.5em}
\subsection{Hardware setup and experimental method}
\label{section:hw}
In our experiments, we run $AT_{Advt}$ and $AT_{Scan}$ on a desktop with AMD Ryzen 9 5900X, 64~GB RAM, and Intel Wi-Fi 6 AX200 network card that enables Bluetooth~5.2. We install  \textit{victim's app} on five smartphones that run the original stock Android-based OS from their manufactures. Table~\ref{tab:devices} reports the configurations for these mobile devices in terms of release year, OS, Bluetooth version supported, and the selected BLE advertising method.
\begingroup
\setlength{\tabcolsep}{4pt} 
\renewcommand{\arraystretch}{0.9}
\begin{table}[!htbp]
	\vspace{-.5em}
	\centering
	\caption{Configurations of victim mobile devices.}
	\vspace{-.65em}
	\label{tab:devices}
	\begin{tabular}{l c c c c c }
		\hline
		Device Model & Release Date & Android Ver.& Operating System & Bluetooth Ver. \\
		\hline
		Oneplus6    & 2018.05      & 10              & H2OS 10.0.11     & 5.0      \\
		Oneplus8    & 2020.04      & 11              & H2OS 11.0.13     & 5.1      \\
		OppoReno4   & 2020.06      & 11              & ColorOS 11      & 5.1     \\
		Redmi10xpro & 2020.05      & 11              & MIUI 12.5.4      & 5.1      \\
		VivoiQooZ1  & 2020.05      & 11              & OriginOS 1.0     & 5.0      \\
		\hline
	\end{tabular}
	\vspace{-1.6em}
\end{table}
\endgroup
\par
We primarily test the performance of our attack by varying the parameter~$t$~(i.e., time between victim’s successive advertisements).
\riccardo{We use a randomly generated text~\cite{LoremIpsum} for the payload to be transmitted.}
\removed{ with a fixed length of 1236~bytes. \textit{Victim's app} divides the payload into 103 advertisements.}
We set $R=3$ and $t=[1, 2, 3]$ seconds while we set the maximum size $z$ of segments according to BLE technology used. For each mobile device, we repeat our experiments three time for each value of $t$.
Our experiment settings (i.e., $R=3$) enable three transmissions of all the payload \textit{packets}, thus, $AT_{Scan}$ can receive duplicate \textit{packets} twice. It is worth mentioning that we exclude duplicate \textit{packets} to evaluate the effective performance of our attack.
We stop few seconds after \textit{victim's app} transmits the last \textit{packet} in sequence. \riccardo{We evaluate the performance according to a thorough set of metrics, i.e., data transfer rate, \textit{packet} loss, \textit{packet} inter-arrival time, total transmission time, and percentage of payload received over time.}
\vspace{-.75em}
\subsection{Experimental results - Legacy advertisements}
\label{section:resultsLegacy}
\begin{figure}[ht]
\vspace{-.25em}
\centering
\subfigure
{
	\includegraphics[trim = 13mm 68mm 5mm 2mm, clip, width=0.65\columnwidth]{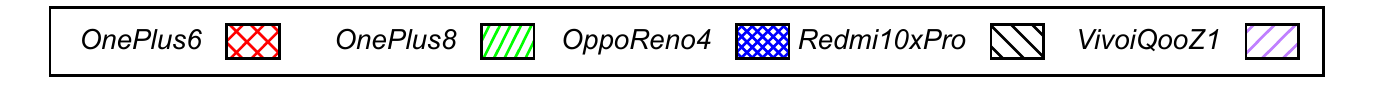}}
\\\vspace{-.5em}
\addtocounter{subfigure}{-1}
\subfigure[Average data transfer rate.]
{ \label{fig:avg_data_rate}
	\includegraphics[trim = 2mm 2mm 5mm 15mm, clip, width=0.45\columnwidth]{images/graphs/3TRS_avg_data_rate_min_max.pdf}}
\subfigure[Average \textit{packet} loss.]
{ \label{fig:avg_data_loss}
	\includegraphics[trim = 2mm 2mm 5mm 15mm, clip, width=0.45\columnwidth]{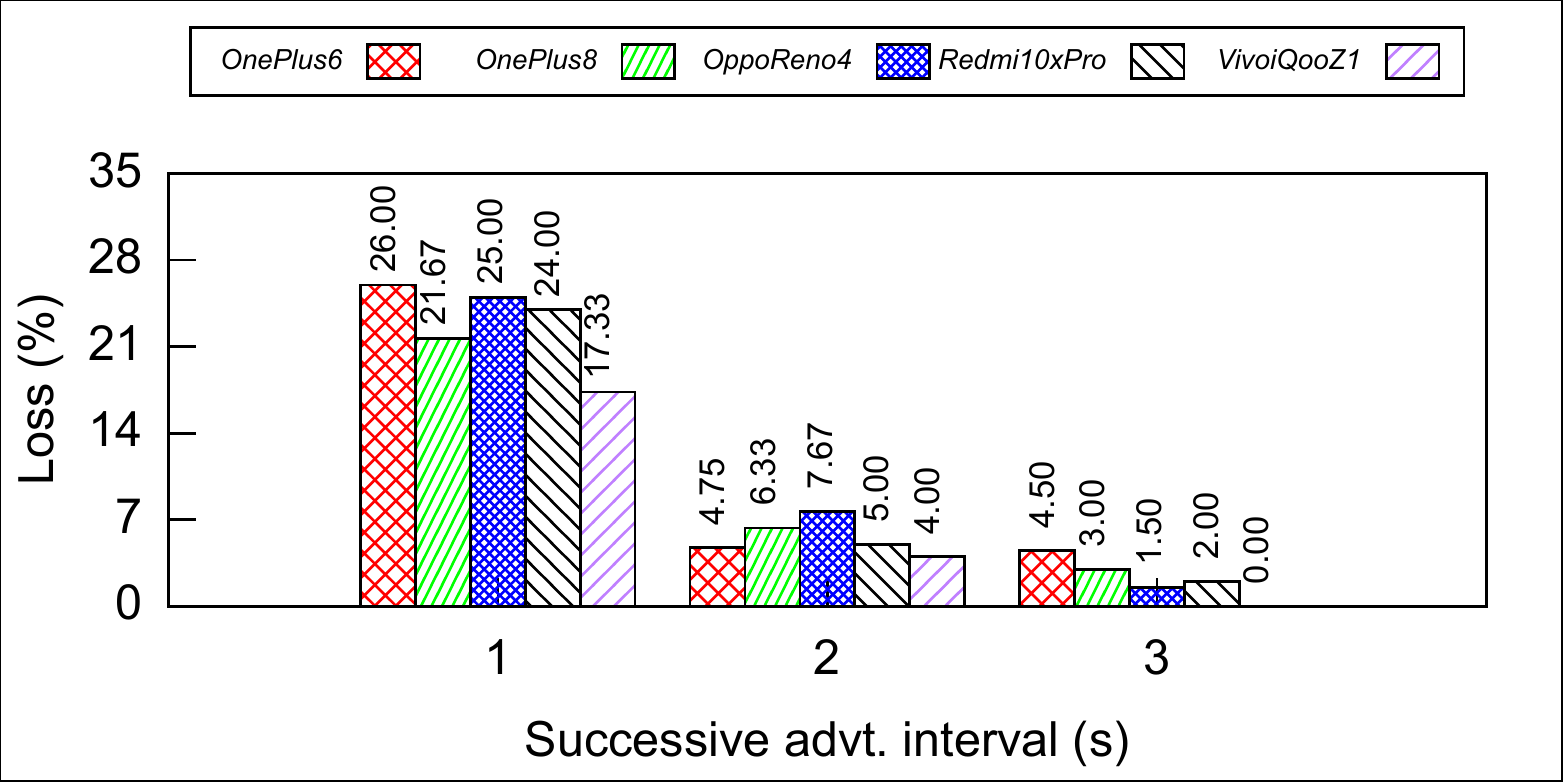}
}
\\
\subfigure[Average \textit{packet} inter-arrival rate.]
{ \label{fig:avg_pkt_interarrival_rate}
	\includegraphics[trim = 2mm 2mm 5mm 15mm, clip, width=0.45\columnwidth]{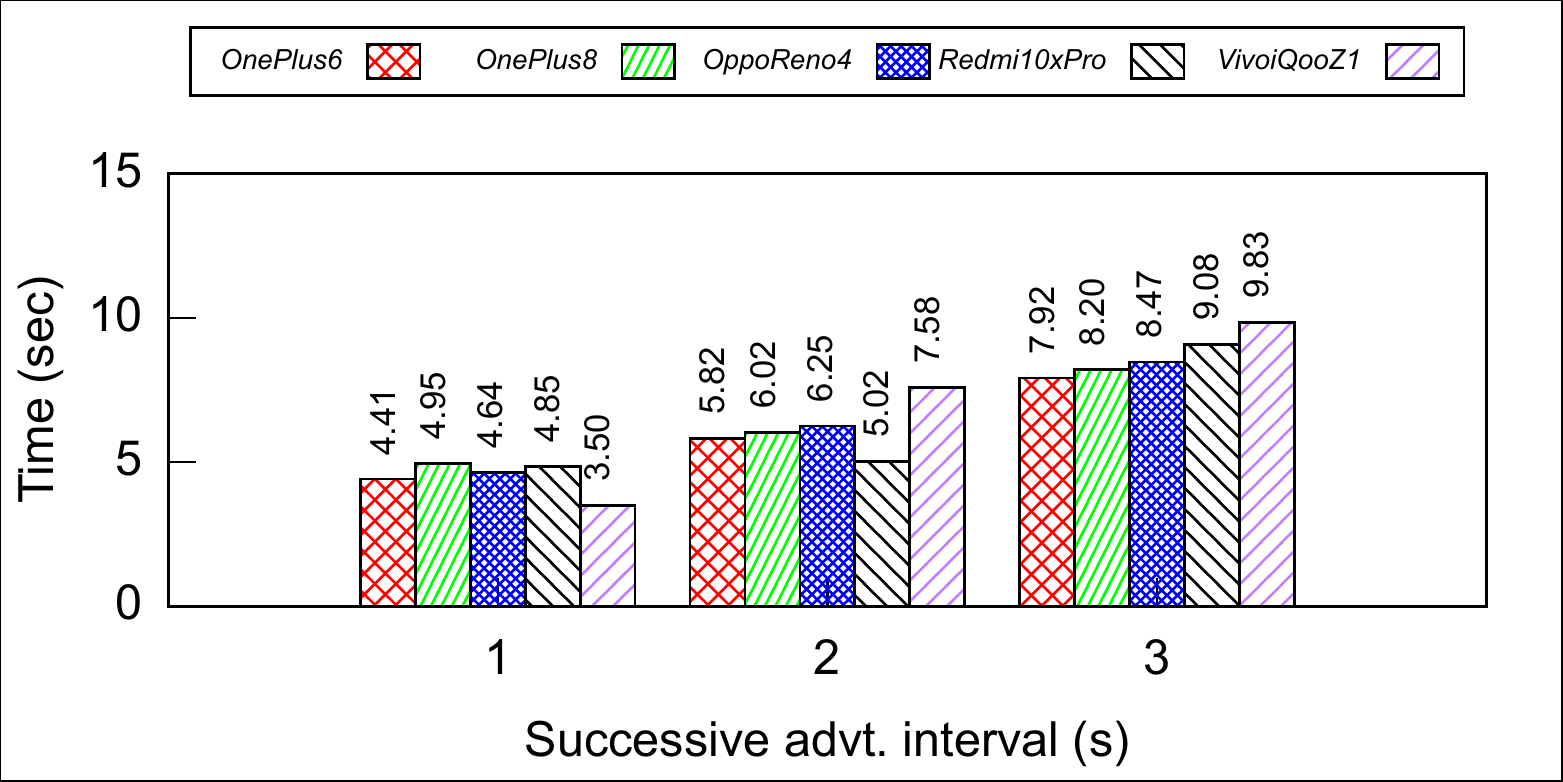}}
\subfigure[Average transmission time.]
{ \label{fig:avg_trans_time}
	\includegraphics[trim = 2mm 2mm 5mm 15mm, clip, width=0.45\columnwidth]{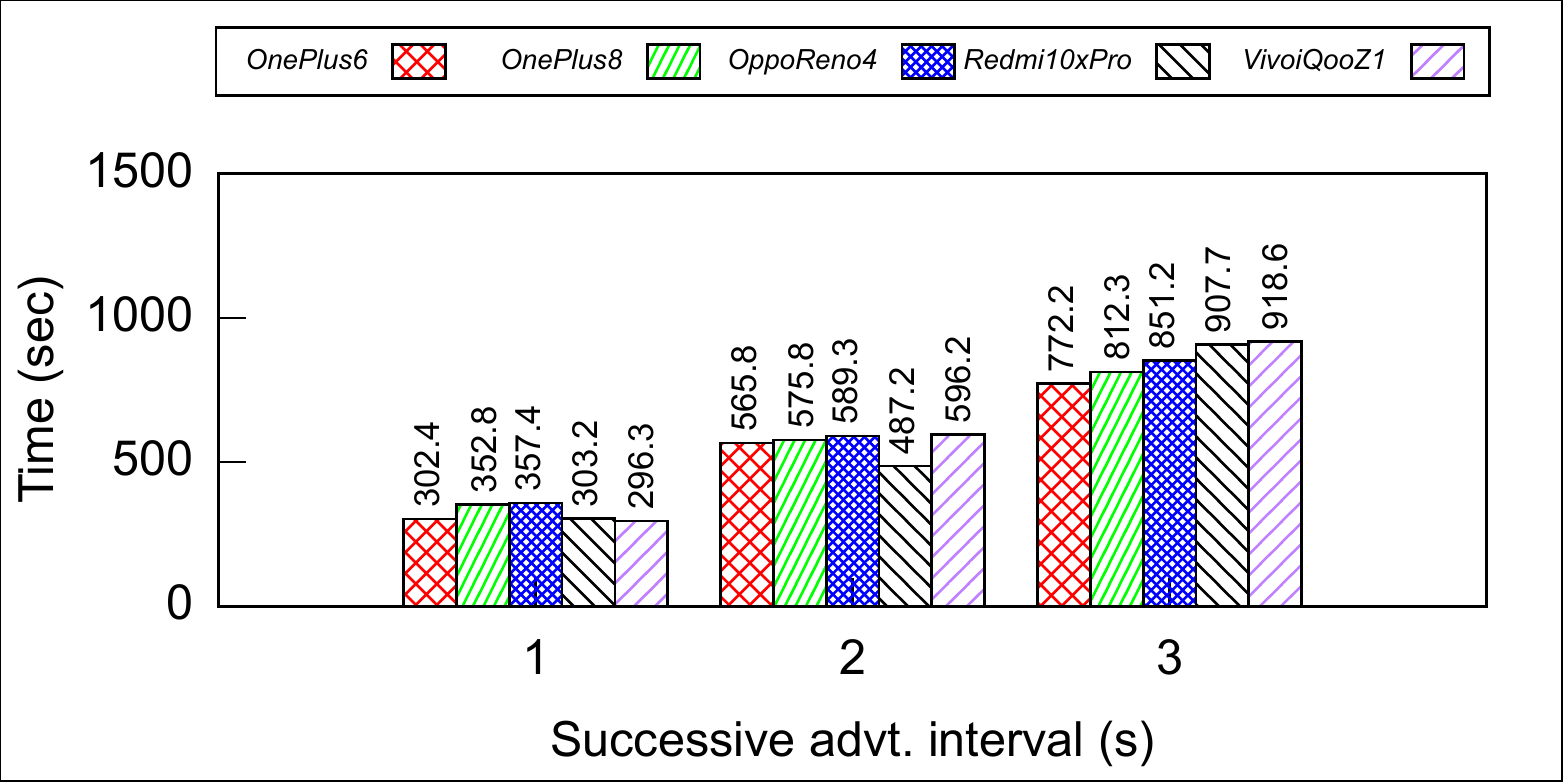}}
\vspace{-.5em}
\caption{Our attack's performance (duplicates excluded) \riccardo{for BLE legacy}.}
\label{fig:results}
\vspace{-1.25em}
\end{figure}

\riccardo{Considering BLE legacy advertising, it allows to (i)~cover a wider range of Bluetooth-enabled devices, and (ii)~show the lower bounds for our attack.} \riccardo{In these experiments, we transmit a payload with a fixed length of $1236$~bytes, which is divided by \textit{victim's app} into a total of $103$ advertisements (i.e., $z=12$~bytes).} 
We report the evaluation results in \figurename{~\ref{fig:results}}. 
In \figurename~\ref{fig:avg_data_rate}, we report the average data transfer rate for the three values of $t$. We can notice that the overall data transfer rate with $t=1$ second is about $3$~bytes/sec while it is reduced to half for $t=3$ (i.e., around $1.5$~bytes/sec). While we achieve a higher transfer rate with $t=1$, we also have a higher percentage of \textit{packet} loss as reported in \figurename{~\ref{fig:avg_data_loss}}. However, the percentage of lost \textit{packets} is drastically reduced by setting $t=2$ and $t=3$, i.e., around $5.5\%$ and $2.2\%$ on average, respectively. In terms of time, we can observe that both the average \textit{packet} inter-arrival time (in \figurename~\ref{fig:avg_pkt_interarrival_rate}) and the average time for three full payload ($R=3$) transmissions (in \figurename~\ref{fig:avg_trans_time}) increase with the value of $t$; while it remains stable among the different device models. In light of these results, we can argue that $t=2$ is a reasonable trade-off between data transfer rate, limited \textit{packet} loss, and total transmission time.
\par
\begin{figure}[ht]
\vspace{-1.2em}
\centering
\subfigure[For $t=1s$.]
{ \label{fig:totpkttimeint1}
	\includegraphics[trim = 2mm 0mm 2mm 2mm, clip, width=0.3\columnwidth]{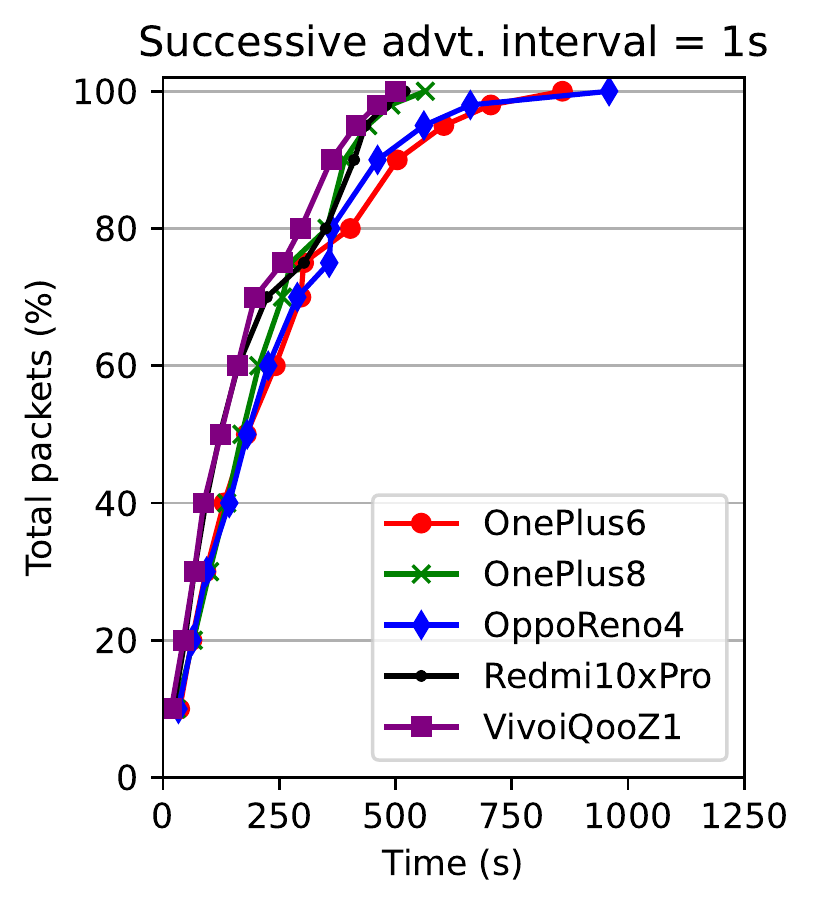}
}\hspace{-2mm}
\subfigure[For $t=2s$.]
{ \label{fig:totpkttimeint2}
	\includegraphics[trim = 2mm 0mm 2mm 2mm, clip, width=0.3\columnwidth]{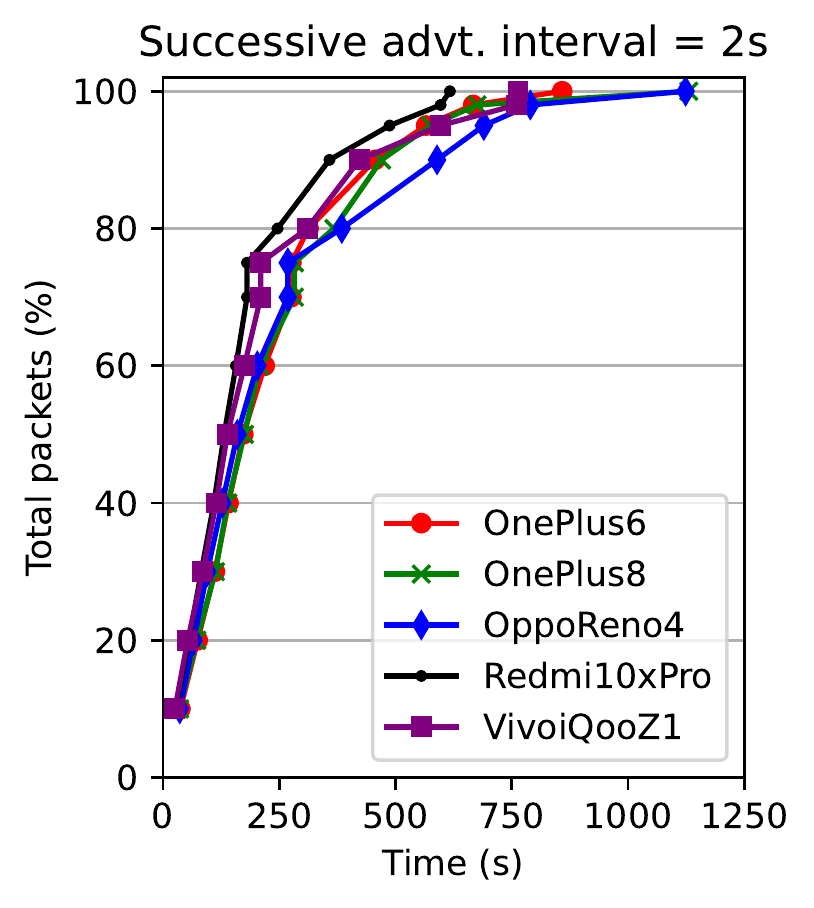}
}\hspace{-2mm}
\subfigure[For $t=3s$.]
{ \label{fig:totpkttimeint3}
	\includegraphics[trim = 2mm 0mm 2mm 2mm, clip, width=0.3\columnwidth]{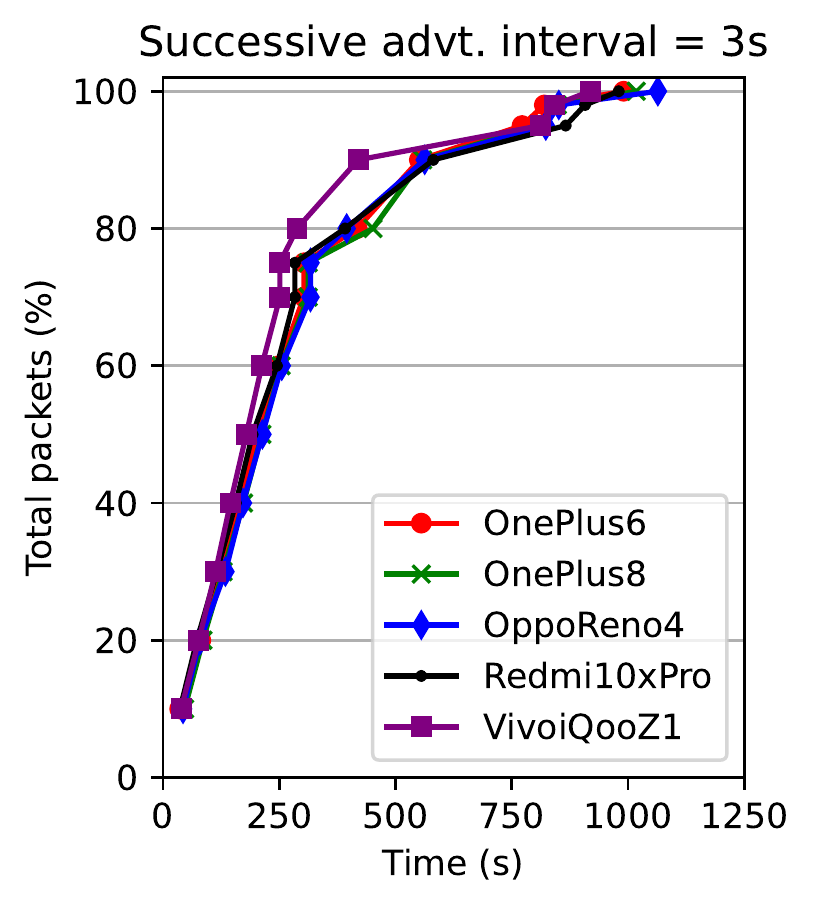}
}
\vspace{-.5em}
\caption{Total \textit{packets} (\%) received (duplicates excluded) \riccardo{over time for BLE legacy.}}
\label{fig:totpkttime}
\vspace{-1.5em}
\end{figure}

As a further analysis, we report in \figurename~\ref{fig:totpkttime} the percentage of received unique \textit{packets} over time. Differently from the previous experiments, here we keep retransmitting the entire sequence of payload \textit{packets} (i.e., $R=\infty$) until a timeout~($T$) at $1250$ seconds.
As a confirmation of our previous results, we receive on average $80\%$ of the total \textit{packets} in $320$~seconds, $93\%$~\textit{packets} in $520$~seconds, and $98\%$~\textit{packets} in $850$~seconds, for $t=1$, $2$, and $3$, respectively. This analysis also highlights that we receive the majority of \textit{packets} (i.e., around $80\%$) within the first $320$~seconds independently from the value of $t$. The remaining $20\%$ \textit{packets} suffer longer transmission time primarily due to blind retransmission of the entire \textit{packet} sequence, augmented by natural transmission losses. As a possible strategy to avoid such a situation, an attacker can set an optimal transmission timeout, and then request retransmission of only missing \textit{packets}.
\vspace{-1.65em}
\subsection{\ankit{Experimental results - Extended advertisements}}
\label{section:resultsExtended}
\begin{figure}[!htbp]
\vspace{-.5em}
\centering
\subfigure
{
	\includegraphics[trim = 13mm 68mm 5mm 2mm, clip, width=0.65\columnwidth]{images/graphs/3TRS_avg_data_rate_min_max.pdf}}
\\\vspace{-.5em}
\addtocounter{subfigure}{-1}
\subfigure[Average data transfer rate.]
{ \label{fig:BLE5avg_data_rate}
	\includegraphics[trim = 2mm 2mm 5mm 15mm, clip, width=0.45\columnwidth]{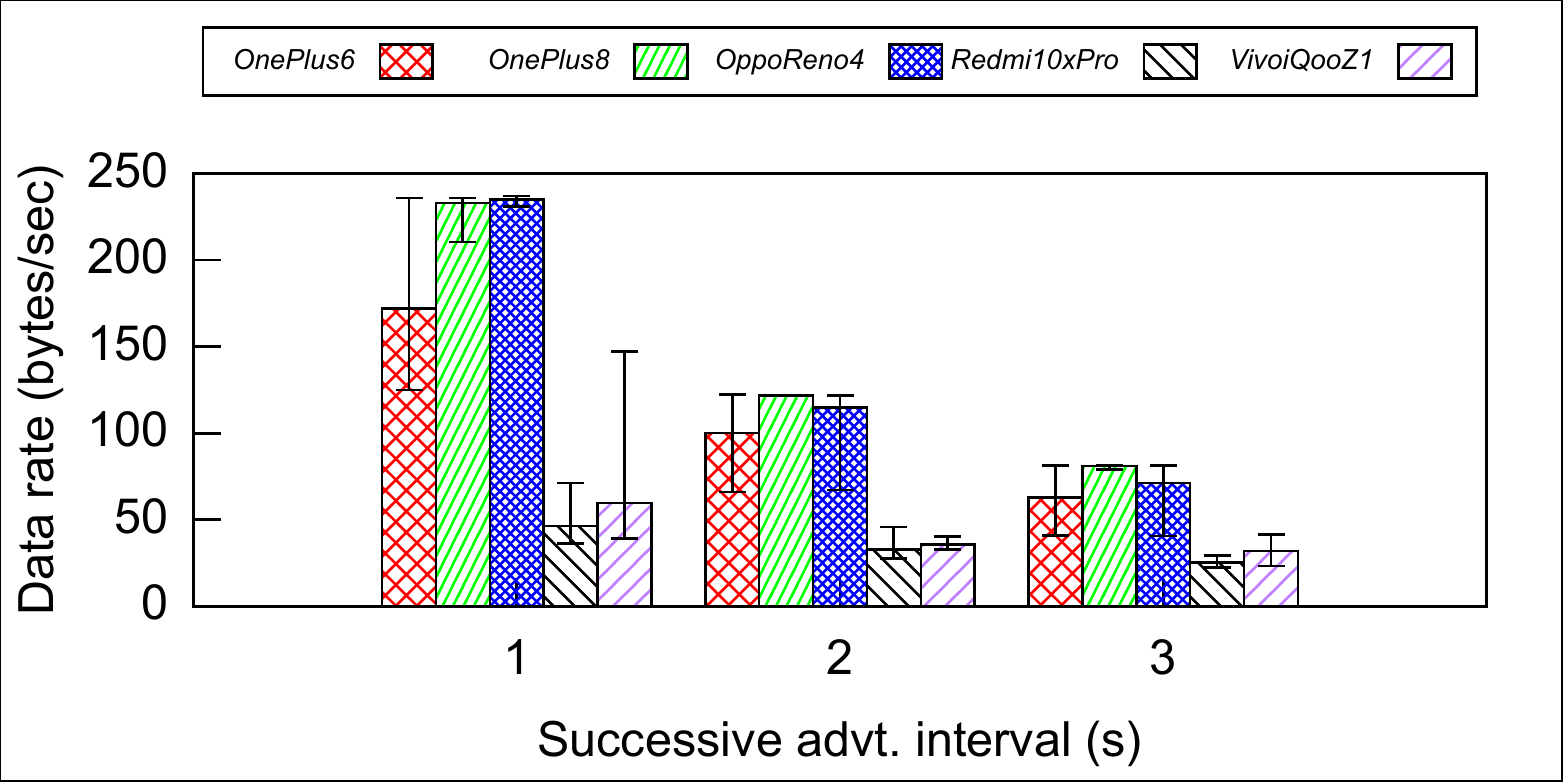}}
\subfigure[Average \textit{packet} loss.]
{ \label{fig:BLE5avg_data_loss}
	\includegraphics[trim = 2mm 2mm 5mm 15mm, clip, width=0.45\columnwidth]{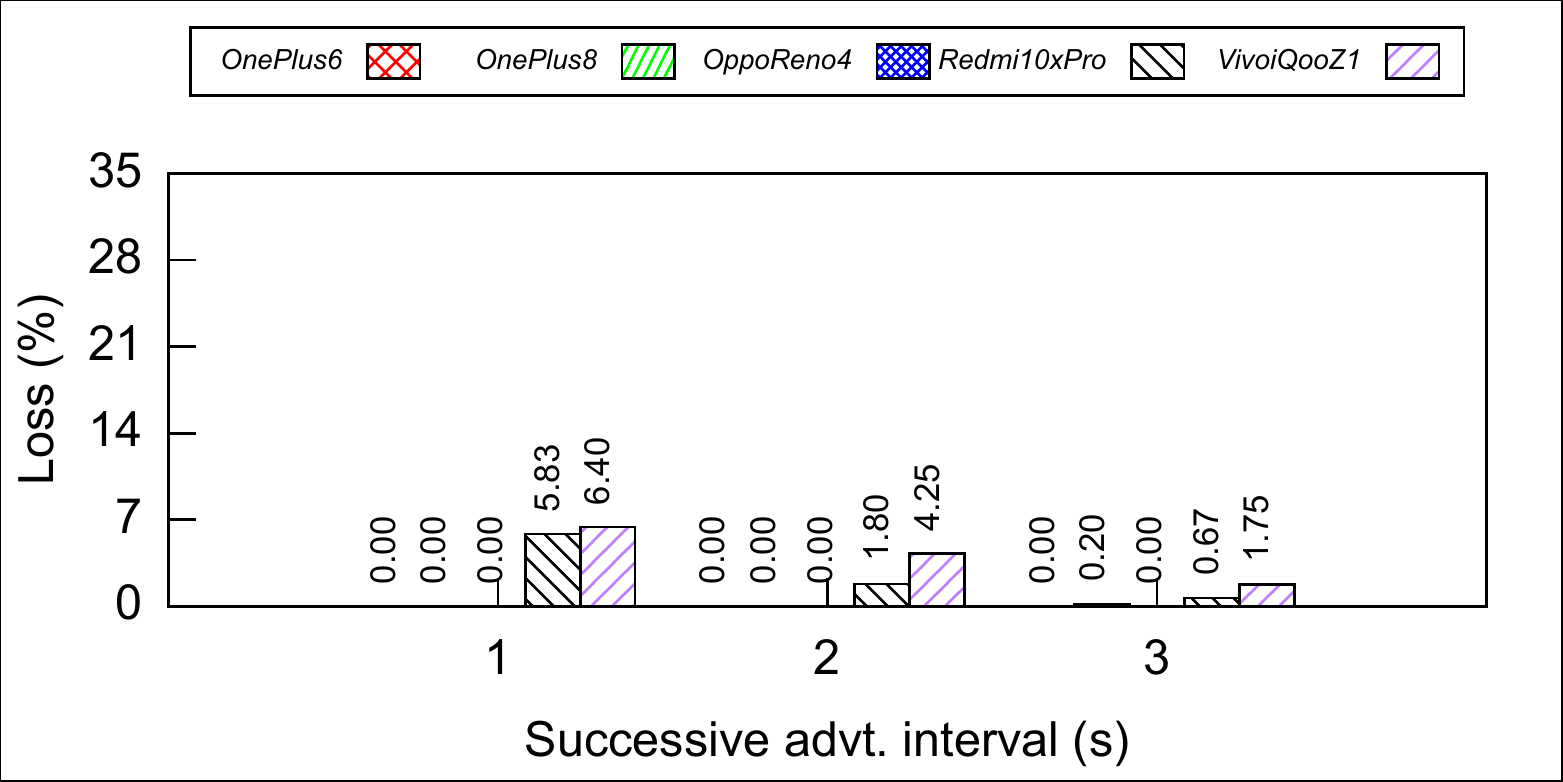}
}
\\
\subfigure[Average \textit{packet} inter-arrival rate.]
{ \label{fig:BLE5avg_pkt_interarrival_rate}
	\includegraphics[trim = 2mm 2mm 5mm 15mm, clip, width=0.45\columnwidth]{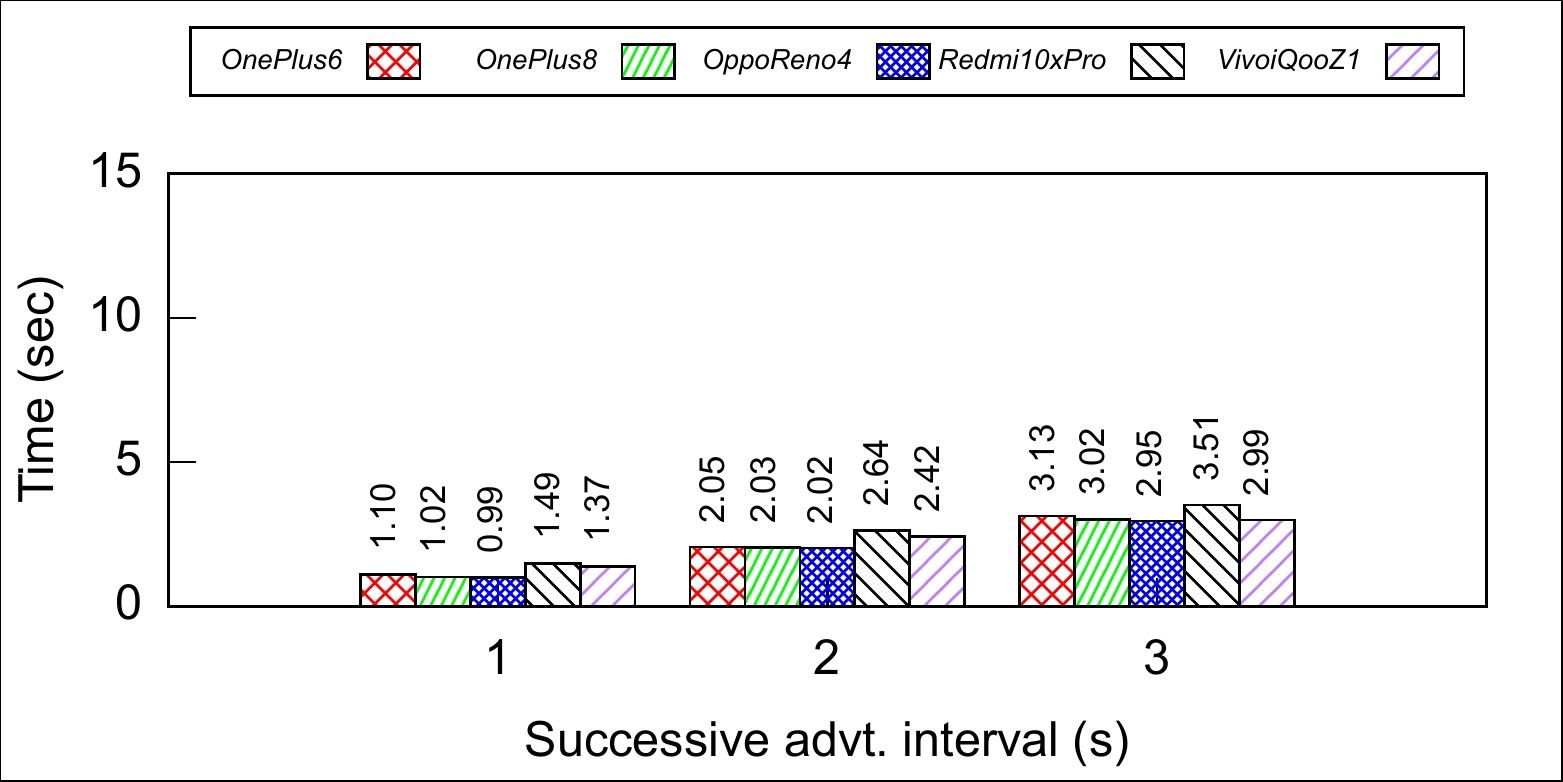}}
\subfigure[Average transmission time.]
{ \label{fig:BLE5avg_trans_time}
	\includegraphics[trim = 2mm 2mm 5mm 15mm, clip, width=0.45\columnwidth]{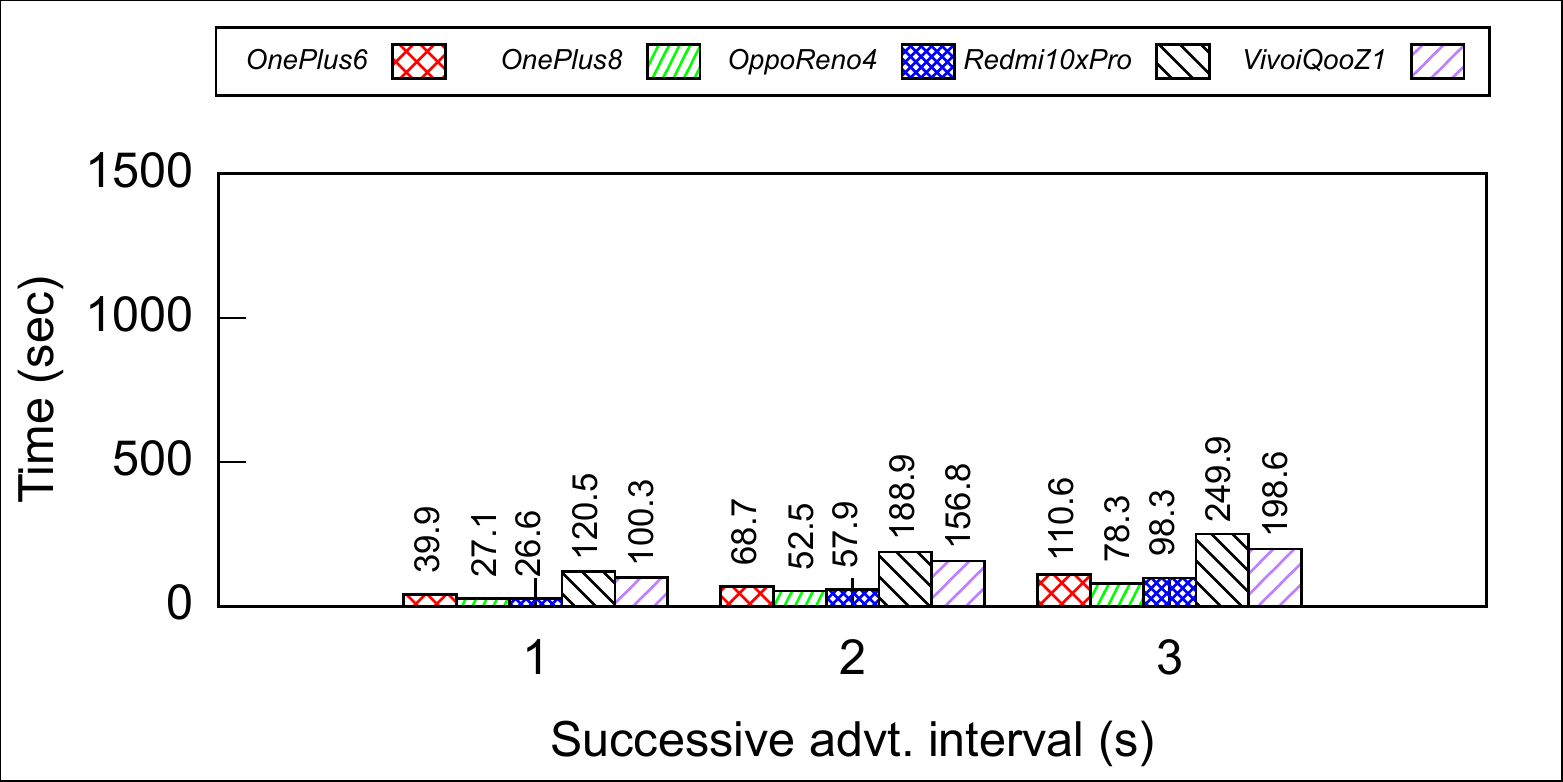}}
\vspace{-.5em}
\caption{\riccardo{Our attack's performance (duplicates excluded) for BLE extended.}}
\label{fig:BLE5results}
\vspace{-1.5em}
\end{figure}

\riccardo{Bluetooth extended advertising~\cite{BL5.2} allows us to improve the data transfer rate for our attack compared to legacy advertisements~\cite{BL4.2}. We set the maximum size~$z$ of segments according to the Maximum Advertising Data Length~(MADL) supported by the considered devices and taking relevant extended headers into account.} 
\ankit{We could set maximum $z=237$~bytes for \textit{Group~A} devices (i.e., Oneplus6, Oneplus8, and OppoReno4) and maximum $z=170$~bytes for \textit{Group~B} devices (i.e., Redmi10xpro and VivoiQooZ1). In these experiments, we transmit a fixed length payload of $6180$~bytes~(5 times of payload used in BLE legacy~experiments), which is divided by \textit{victim's app} into a total of $37$ and $27$ advertisements for $z=170$ and $z=237$ bytes, respectively. \figurename{~\ref{fig:BLE5results}} reports our results.}
\vspace{-.5em}
\par
\riccardo{As expected MADL, and thus $z$, has a significant impact on the overall performance. As reported in \figurename~\ref{fig:BLE5avg_data_rate}, we achieve an average data transfer rate of up to $236$ and $60$ bytes/sec for devices in \textit{Group~A} and \textit{Group~B}, respectively. Compared to legacy advertising, it corresponds to an improvement of $78$ times for \textit{Group~A} and $20$ times for \textit{Group~B}.  
In \figurename{~\ref{fig:BLE5avg_data_loss}}, we notice that \textit{Group~A} experience almost no \textit{packet} loss while \textit{Group~B} suffers at most $7\%$ \textit{packet} loss, which is also significantly lower than the one from legacy advertising. 
Considering the time-related metrics, average \textit{packet} inter-arrival rate~(cf. \figurename~\ref{fig:BLE5avg_pkt_interarrival_rate}) almost coincides with interval $t$, and it determines the total transmission time~(cf. \figurename~\ref{fig:BLE5avg_trans_time}) also according to transfer rate of devices in \textit{Group~A} and \textit{Group~B}. }
\par
\begin{figure}[!htbp]
\vspace{-1em}
\centering
\subfigure[For $t=1s$.]
{ \label{fig:BLE5totpkttimeint1}
	\includegraphics[trim = 2mm 0mm 2mm 2mm, clip, width=0.3\columnwidth]{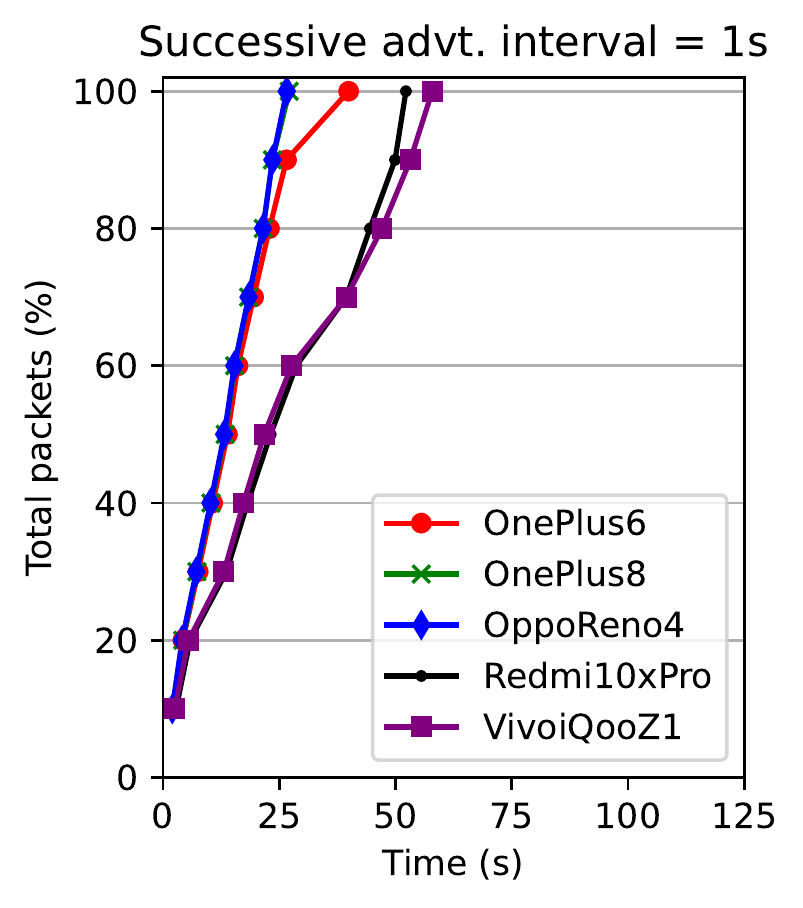}
}\hspace{-2mm}
\subfigure[For $t=2s$.]
{ \label{fig:BLE5totpkttimeint2}
	\includegraphics[trim = 2mm 0mm 2mm 2mm, clip, width=0.3\columnwidth]{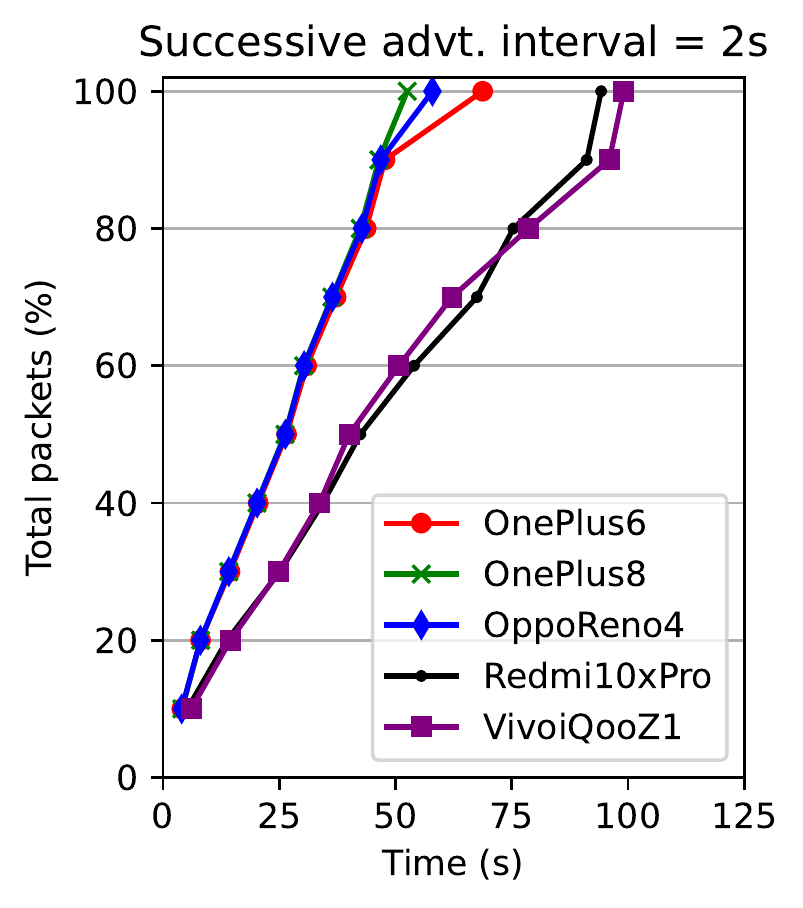}
}\hspace{-2mm}
\subfigure[For $t=3s$.]
{ \label{fig:BLE5totpkttimeint3}
	\includegraphics[trim = 2mm 0mm 2mm 2mm, clip, width=0.3\columnwidth]{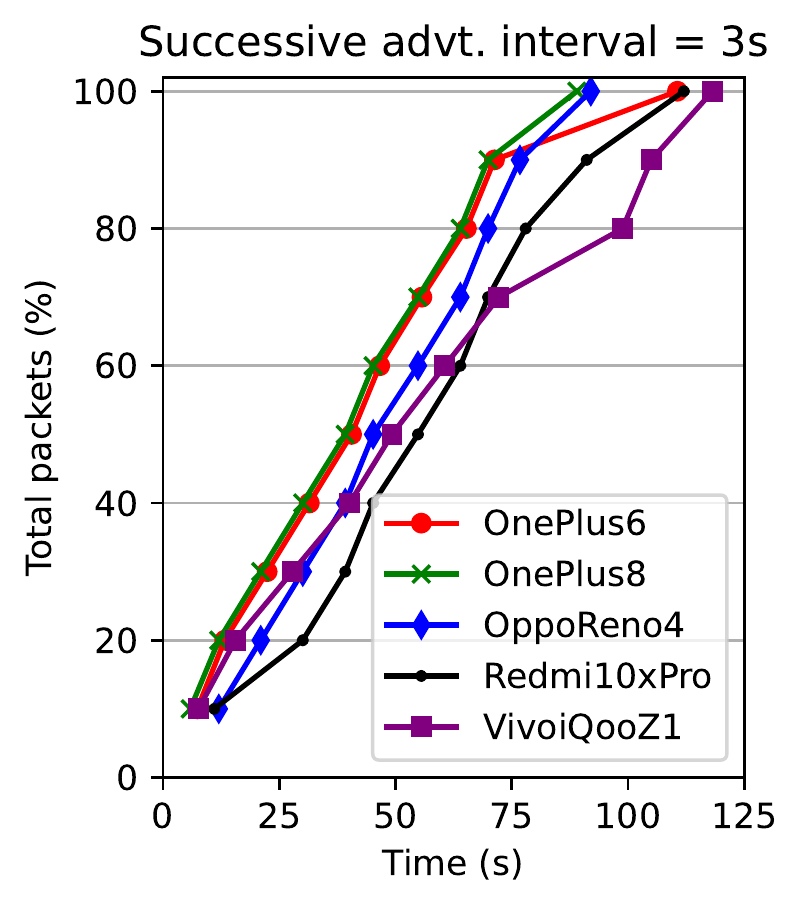}
}
\vspace{-.75em}
\caption{\riccardo{Total \textit{packets} (\%) received (duplicates excluded) over time for BLE extended.}}
\label{fig:BLE5totpkttime}
\vspace{-1em}
\end{figure}

\riccardo{
\figurename{~\ref{fig:BLE5totpkttime}} reports the percentage of unique \textit{packets} received over time for extended advertising with $R=\infty$ and timeout~$T=125$~seconds \ankit{($1/10^{th}$ of $T$ set in BLE legacy experiments)}. In \textit{Group~A}, we receive the $90\%$ of \textit{packets} within about $25$, $50$, and  $75$~seconds for $t=1$, $2$, and $3$, respectively. Considering the same values of $t$, we receive all \textit{packets} within about $60$, $100$, and  $120$~seconds in \textit{Group~B}. Comparing with the results of the same study on legacy advertising~(cf. \figurename{~\ref{fig:totpkttime}}), we argue that extended advertising enables a more reliable transmission due to low \textit{packet} loss rate and reasonable total transmission time. Therefore, we do not need to apply the strategy based on timeout and selective packet retransmissions discussed in Section~\ref{section:resultsLegacy}. 
} 

\vspace{-.75em}
\section{Discussion}
\label{section:discussion}
\vspace{-.5em}
\riccardo{Here, we discuss the potential application of our attack in the real world, methods to boost the transfer speed for exfiltration of large files, constraints related to the Android OS versions, and possible countermeasures against our attack.}
\subsection{Attack scenarios}
\label{section:wayToUse}
We use our attack with extended advertisements to exfiltrate sensitive information from a victim's device. On a OnePlus6 device, we configure our \textit{victim's app} to exfiltrate several types of sensitive information as use cases. We quantify both the size of such information and the time required for the exfiltration using BLE extended. It is important to note that the following use cases do not require additional permissions unless explicitly specified.

\textit{$\sbullet[.75]$ Get device build information:} 
An installed app can access information about the current build (e.g., OS Version, API level, security patch level) through Build~\cite{G_Build} class of Android SDK. 
An adversary can use our attack to gain access to such information and exploit a vulnerability specific to that build. \riccardo{Due to its small size, our attack can exfiltrate such information in two seconds at most.}

\textit{$\sbullet[.75]$ Get list of installed apps:}
Any installed app can access the list of apps currently installed on the device using PackageManager~\cite{G_PackageManager} class. 
An adversary can use such a list of installed apps to exploit known vulnerabilities in the installed packages or even to predict user traits from the list of installed apps~\cite{Seneviratne2014apps}. \riccardo{On OnePlus6, we successfully exfiltrate a list of $496$ apps installed (i.e., package names; comprising $3.9$KB in compressed form) in less than $17$ seconds.}

\textit{$\sbullet[.75]$ Extract information accessible to \textit{victim's app}:} \textit{Victim's app} disguises as a benign app, which may require some permissions depending on the context. E.g., \textit{victim's app} pretending to be a fitness app might ask for storage permission. Our attack can exfiltrate such context-specific information accessible to \textit{victim's app}. \riccardo{E.g., we successfully exfiltrate one thousand contacts (i.e., full name, email, and phone number; comprising $17.1$KB in compressed form) in about $75$~seconds. As another example, we exfiltrate one thousand calendar events (in an ICS file; comprising $65.3$KB in compressed form) in about $285$~seconds.}
Moreover, an attacker can leverage our mapping of $UUID_{V}$ to $ID_{V}$~(cf. Section{~\ref{section:thecoreIdea}}) to bypass Bluetooth's address randomization defense that prevents device tracking~\cite{fawaz2016protecting}.

\textit{$\sbullet[.75]$ Deliver malicious payload:}
Our attack opens an avenue to deliver malicious payload to victim devices. The attacker can broadcast a malicious payload through a series of advertisements  from its $UUID_{A}$; this process is similar to our regular attack, only the roles of sender and receiver have reversed. Alternatively, the attacker can deliver larger payload via a WiFi network connection by using socket programming, where our attack is used as C\&C to steer WiFi/LocalOnlyHotspot connections (cf. Section{~\ref{section:fastTransfer}}). Once the malicious payload is delivered, InMemoryDexClassLoader~\cite{G_InMemoryDexClassLoader} can execute it via ByteBuffer~\cite{G_ByteBuffer}. Since all the components are in the buffer, we do not need storage permissions.

\subsection{Boosting data transfer speed}
\label{section:fastTransfer}
\ankit{Despite the improved transfer rates enabled by extended advertisements, our attack may not be suitable to exfiltrate large files, e.g., high-resolution photos. So, we investigate whether our proposed attack can be further strengthened in terms of data rate.} 
\riccardo{We identify two viable solutions to boost the data transfer speed using other faster wireless communication channels~(e.g., WiFi), where our attack is leveraged as C\&C to enable the alternative wireless channel.} 
\begin{enumerate}
\item \textit{Connecting to an attacker-controlled WiFi network:}
Android 9 and below allow us to toggle WiFi connection without user's permission and to connect to a particular WiFi Access Point~(AP) by specifying its SSID and password. The attacker can create an AP and send commands to the victim device~(via our attack) asking it to connect to the AP. Then, this connection can be used for fast data extraction, e.g., over a peer-to-peer WiFi file sharing system. 
\lstlistingname~\ref{wifiSettings} in \appendixname{~\ref{appendix:codeSnippets}} shows turning WiFi on and connecting to a particular SSID. Additional permissions required here are \texttt{ACCESS\_NETWORK\_STATE}, \texttt{ACCESS\_WIFI\_STATE}, \texttt{CHANGE\_WIFI\_STATE}, and \texttt{INTERNET}~(only to open network sockets). All these permissions are normal permissions. 

\item \textit{Using startLocalOnlyHotspot:} Android 10+ may restrict the above mentioned method of connecting to an arbitrary WiFi AP because the decision~\cite{G_WiFiSuggest} to select/prefer an AP is made by the underlying OS. To overcome this restriction, \textit{victim's app} can create a local hotspot~(irrespective of Internet access) using startLocalOnlyHotspot~\cite{G_startLocalOnlyHotspot}
Upon successful creation of hotspot, the reservation object returns SSID, security type, and credentials for connecting to such hotspot. \textit{Victim's app} can pass these credentials to attacker via our attack channel. Then, attacker can exploit this hotspot connection in the same way as attacker-controlled WiFi AP~(discussed above). \lstlistingname~\ref{localHotSpot} in \appendixname{~\ref{appendix:codeSnippets}} demonstrates using startLocalOnlyHotspot. Additional permissions required here are \texttt{CHANGE\_WIFI\_STATE}, \texttt{INTERNET}~(only to open network sockets), and \texttt{ACCESS\_FINE\_LOCATION}; the first two permissions are normal permissions while the last one is already available with \textit{victim's app}.
\end{enumerate}

\subsection{Android version-specific requirements}
\label{section:and10+}
Our attack require Bluetooth~(i.e., \texttt{BLUETOOTH} and \texttt{BLUETOOTH\_ADMIN}) and location~(i.e., \texttt{ACCESS\_FINE\_LOCATION}) permissions. Both Bluetooth-related permissions are normal and will remain the same for Android 12 (API level 31). Location permission is classified as dangerous, hence \textit{victim's app} disguises as benign app to obtain this permission. Till Android 9, location permission is only obtained, but location service is not required to be turned on. But in Android 10+, location service needs to be turned on to get scanning results. It can be seen as a limitation, and to bypass it the attacker must disguise \textit{victim's app} as a genuine app that needs location service to be on~(e.g., as contact tracing app). 

\subsection{Countermeasures}
\label{section:countermeasures}
Our proposed attack primarily exploits Bluetooth channel. According to the Android permission documentation~\cite{AndroidManifestPermission},  \texttt{BLUETOOTH} and \texttt{BLUETOOTH\_ADMIN} are normal permissions and will be the same for Android 12 (API level 31). Although starting Android 12, \texttt{BLUETOOTH\_ADVERTISE} has become a dangerous permission, we recommend that \texttt{BLUETOOTH} permission itself is made a dangerous permission so that the user is notified if an application accesses Bluetooth in any form. Permissions can be obtained from an average user by using apt pretexts~\cite{apt_pretext1, apt_pretext2}. Hence, our attack remains valid even with the new permissions introduced in Android 12 (API level 31). \riccardo{We propose several concrete OS-based, and thus user-independent, countermeasures to limit the capabilities of our attack:
\begin{enumerate}
\item The OS should inherently prevent continuous advertising by apps and/or increase the time interval between advertisements. 
\item As advertisements are mainly used to broadcast connection parameters and preferences, it is reasonable to expect that the advertisement content - differently from our attack - would not change frequently. \ankit{So, the OS can impose a limit on the frequency of such changes in} consecutive advertisements.
\item The content of advertisements should undertake strict control. The OS can employ semantic checks or taint analysis to identify anomalous content that could indicate a data exfiltration attempt.
\item The OS can also restrict the content of advertisements to a list of predetermined values. Despite an attacker can still use such values as a basis to encode the to-be-transferred information (e.g., value\#1 = `0' and value\#2 = `1'), such a measure will drastically reduce the throughput of our attack.
\end{enumerate}} 

\section{Conclusion}
\label{section:conclusion}
BLE extends BT stack with limited energy requirements and provide convenient functionalities, making it suitable for many industrial and consumer applications. Among such functionalities, BLE advertisements ease the discovery of other in-range devices. In this paper, we proposed an attack that exploits BLE advertisements' Service Data field to establish a communication medium between unpaired devices. We discussed how an attacker can leverage this communication channel as a building block for data exfiltration from a device and cater to even more dangerous attacks. While our proof-of-concept implementation considers legacy BLE advertising to cover the widest range of BLE devices, its data ~~ transfer rate can be increased by using extended advertisements in BLE~5.0. Therefore, we argue that misuse of BLE advertisements poses a significant security threat, which can be limited by adopting our proposed countermeasures.

\bibliographystyle{splncs04}
\bibliography{bib}
\balance
\vfil

\begin{subappendices}
\renewcommand{\thesection}{\Alph{section}}
\section{Code snippets}
\label{appendix:codeSnippets}
Here, we report the code required for: 
(i)~advertisement manipulation by $AT_{Advt}$ in \lstlistingname~\ref{lst:advertiserModel}, 
(ii)~\textit{victim's app} scanner mode configuration in \lstlistingname~\ref{bluetoothLeScannerSettings},
(iii)~\textit{victim's app} advertiser mode configuration in \lstlistingname~\ref{setAdvertiseSettingsAdvertiseCallback},
(iv)~turning WiFi on and connecting to a specific SSID in \lstlistingname~\ref{wifiSettings}, and 
(v)~starting startLocalOnlyHotspot and getting credentials for the hotspot in \lstlistingname~\ref{localHotSpot}.

\begin{lstlisting}[xleftmargin=0.0ex, xrightmargin=-0.0ex, caption={DataSection content manipulation $AT_{Advt}$},captionpos=b, ,label=lst:advertiserModel]
using Windows.Devices.Bluetooth.Advertisement;
private static BluetoothLEAdvertisementPublisher blePublisher = new BluetoothLEAdvertisementPublisher();
private static void sendCommandAdvt(Guid advertiserUUID, Byte[] command, String VictimId){
List<byte> data2send = new List<byte>();
data2send.AddRange(advertiserUUID.ToByteArray())
data2send.AddRange(command)
if (command!= (byte)0){
	data2send.AddRange(Encoding.UTF8.GetBytes(VictimId))}
IDataWriter dataWriter = new DataWriter();
dataWriter.WriteBytes(data2send.ToArray());
IBuffer buffer = dataWriter.DetachBuffer();
BluetoothLEAdvertisementDataSection dataSection = new BluetoothLEAdvertisementDataSection(BitConverter.GetBytes(33)[0], buffer);
blePublisher.Advertisement.DataSections.Clear(); //remove default content of ServiceData
blePublisher.Advertisement.DataSections.Add(dataSection); //add command
blePublisher.Start();
}
\end{lstlisting}

\begin{lstlisting}[xleftmargin=0.0ex, xrightmargin=-0.0ex, caption={\textit{Victim's app's} BluetoothLeScanner},captionpos=b, ,label=bluetoothLeScannerSettings]
BluetoothAdapter bluetoothAdapter = BluetoothAdapter.getDefaultAdapter();
ScanSettings scanSettings = new ScanSettings.Builder()
.setScanMode(ScanSettings.SCAN_MODE_BALANCED)
.setCallbackType(ScanSettings.CALLBACK_TYPE_ALL_MATCHES)
.setMatchMode(ScanSettings.MATCH_MODE_AGGRESSIVE)
.setNumOfMatches(ScanSettings.MATCH_NUM_ONE_ADVERTISEMENT)
.setReportDelay(0L).build();
BluetoothLeScanner bluetoothLeScanner = bluetoothAdapter.getBluetoothLeScanner();
List<ScanFilter> scanFilters = new ArrayList<>();
ScanFilter scanFilter = new ScanFilter.Builder().build();
scanFilters.add(scanFilter);
bluetoothLeScanner.startScan(scanFilters,scanSettings,leScanCallback);
\end{lstlisting}

\begin{lstlisting}[xleftmargin=0.0ex, xrightmargin=-0.0ex, caption={Setting AdvertiseSettings and AdvertiseCallback on \textit{victim's app}},captionpos=b, ,label=setAdvertiseSettingsAdvertiseCallback]
BluetoothLeAdvertiser advertiser = bluetoothAdapter.getBluetoothLeAdvertiser();
AdvertiseSettings settings = new AdvertiseSettings.Builder()
.setAdvertiseMode(AdvertiseSettings.ADVERTISE_MODE_BALANCED)
.setTxPowerLevel(AdvertiseSettings.ADVERTISE_TX_POWER_HIGH)
.setConnectable(false).build();
AdvertiseCallback advertiseCallback = new AdvertiseCallback() {
@Override
public void onStartSuccess(AdvertiseSettings settingsInEffect) {
	super.onStartSuccess(settingsInEffect);
	Log.d("Advertisement","Advertise Started");}
@Override
public void onStartFailure(int errorCode) {
	super.onStartFailure(errorCode);
	Log.d("Advertisement","Advertise error "+errorCode);
}
};
\end{lstlisting}

\begin{lstlisting}[xleftmargin=0.0ex, xrightmargin=-0.0ex, caption={Turning WiFi on and connecting to a specific SSID},captionpos=b, ,label=wifiSettings]
WifiManager wifiManager = (WifiManager) getSystemService(WIFI_SERVICE);
wifiManager.setWifiEnabled(true);
String ssid = "LAPTOP-NTT7FOC3 1338";
WifiConfiguration wifiConfiguration = new WifiConfiguration();
wifiConfiguration.SSID = ssid;
wifiConfiguration.preSharedKey = "08(2aR00";
int netID = wifiManager.addNetwork(wifiConfiguration);
wifiManager.disconnect();
wifiManager.enableNetwork(netID,true);
Log.d("WIFI net ID",String.valueOf(netID));
wifiManager.reconnect();
\end{lstlisting}

\begin{lstlisting}[xleftmargin=0.0ex, xrightmargin=-0.0ex, caption={Starting startLocalOnlyHotspot and getting credentials for the hotspot},captionpos=b, ,label=localHotSpot]
WifiManager wifiManager = (WifiManager) getApplicationContext().getSystemService(Context.WIFI_SERVICE);
wifiManager.startLocalOnlyHotspot(new WifiManager.LocalOnlyHotspotCallback() {
@Override
public void onStarted(WifiManager.LocalOnlyHotspotReservation reservation) {
	super.onStarted(reservation);        Log.d("HOTSPOT",reservation.getWifiConfiguration().toString());
	WifiConfiguration config = reservation.getWifiConfiguration();
	String SSID= config.SSID;
	String password=config.preSharedKey;
}
}, new Handler());
\end{lstlisting}

\end{subappendices}	

\end{document}